\documentclass{emulateapj}

\usepackage{psfig}
\usepackage{amsmath}
\usepackage{amssymb}
\usepackage{graphicx}
\usepackage{appendix}
\usepackage{color}
\usepackage{multirow}
\usepackage[normalem]{ulem}

\newcommand{\lSect}[1]{{\label{sec:#1}}}
\newcommand{\lFig}[1]{{\label{fig:#1}}}
\newcommand{\lEq}[1]{{\label{eq:#1}}}
\newcommand{\lTab}[1]{{\label{tab:#1}}}
\def\gtaprx {\lower .1ex\hbox{\rlap{\raise .6ex\hbox{\hskip .3ex
	{\ifmmode{\scriptscriptstyle >}\else
		{$\scriptscriptstyle >$}\fi}}}
	\kern -.4ex{\ifmmode{\scriptscriptstyle \sim}\else
		{$\scriptscriptstyle\sim$}\fi}}}
\def\ltaprx {\lower .1ex\hbox{\rlap{\raise .6ex\hbox{\hskip .3ex
	{\ifmmode{\scriptscriptstyle <}\else
		{$\scriptscriptstyle <$}\fi}}}
	\kern -.4ex{\ifmmode{\scriptscriptstyle \sim}\else
		{$\scriptscriptstyle\sim$}\fi}}}
\newcommand{\FIGFF}[2]{{\ref{fig:#2}{#1}}}

\newcommand{\FIG}[2]{{Fig.~\FIGFF{#1}{#2}}}
\newcommand{\Fig}[1]{{\FIG{}{#1}}}

\newcommand{\Sectff}[1]{{\ref{sec:#1}}}
\newcommand{\Sect}[1]{{\S\Sectff{#1}}}

\newcommand{\Eqref}[1]{{\ref{eq:#1}}}
\newcommand{\Eqff}[1]{{(\Eqref{#1})}}
\newcommand{\eqff}[1]{{\Eqref{#1}}}

\newcommand{\Eq}[1]{{eq.~\Eqff{#1}}}
\newcommand{\eq}[1]{{equation~\eqff{#1}}}

\newcommand{\Msun}{\ensuremath{\mathrm{M}_\odot}}
\newcommand{\Lsun}{\ensuremath{\mathrm{L}_\odot}}

\newcommand{\Zsun}{\ensuremath{\mathrm{Z}_\odot}}
\newcommand{\Tab}[1]{{Table~\ref{tab:#1}}}

\newcommand{\KEPLER}{\ensuremath{\mathrm{\texttt{KEPLER}}}}

\def\gtaprx {\lower .1ex\hbox{\rlap{\raise .6ex\hbox{\hskip .3ex
	{\ifmmode{\scriptscriptstyle >}\else
		{$\scriptscriptstyle >$}\fi}}}
	\kern -.4ex{\ifmmode{\scriptscriptstyle \sim}\else
		{$\scriptscriptstyle\sim$}\fi}}}
\def\ltaprx {\lower .1ex\hbox{\rlap{\raise .6ex\hbox{\hskip .3ex
	{\ifmmode{\scriptscriptstyle <}\else
		{$\scriptscriptstyle <$}\fi}}}
	\kern -.4ex{\ifmmode{\scriptscriptstyle \sim}\else
		{$\scriptscriptstyle\sim$}\fi}}}

\begin{document}

%\submitted{}
%\accepted{}

\title{The Birth Function for Black Holes and Neutron Stars in Close Binaries}

\author{S.\ E.\ Woosley\altaffilmark{1},
        Tuguldur Sukhbold\altaffilmark{2,3,4},
        and H.-T. Janka\altaffilmark{5}}
\altaffiltext{1}{Department of Astronomy and Astrophysics, University
  of California, Santa Cruz, CA 95064}
\altaffiltext{2}{Department of Astronomy, Ohio State University,
  Columbus, Ohio, 43210}
\altaffiltext{3}{Center for Cosmology and AstroParticle Physics, 
  Ohio State University, Columbus OH 43210}
\altaffiltext{4}{NASA Hubble Fellow}
\altaffiltext{5}{Max-Planck-Institut f{\"u}r Astrophysik, Postfach
  1317, 85741 Garching, Germany}

\begin{abstract} 
  The mass function for black holes and neutron stars at birth
    is explored for mass-losing helium stars. These should resemble,
    more closely than similar studies of single hydrogen-rich stars,
    the results of evolution in close binary systems. The effects of
  varying the mass-loss rate and metallicity are calculated using a
  simple semi-analytic approach to stellar evolution that is tuned to
  reproduce detailed numerical calculations. Though the total fraction
  of black holes made in stellar collapse events varies considerably
  with metallicity, mass-loss rate, and mass cutoff, from 5\% to 30\%,
  the shapes of their birth functions are very similar for all
  reasonable variations in these quantities. Median neutron star
  masses are in the range 1.32 -- 1.37 \Msun \ regardless of
  metallicity. The median black hole mass for solar metallicity is
  typically 8 to 9 \Msun\ if only initial helium cores below 40
  \Msun\ (ZAMS mass less than 80 \Msun) are counted, and 9 --
  13 \Msun, in most cases, if helium cores with initial masses up to
  150 \Msun\ (ZAMS mass less than 300 \Msun) contribute. As long as
  the mass-loss rate as a function of mass exhibits no strong
  non--linearities, the black hole birth function from 15 to 35
  \Msun\ has a slope that depends mostly on the initial mass function
  for main sequence stars.  These findings imply the possibility of
  constraining the initial mass function and the properties of mass
  loss in close binaries using ongoing measurements of gravitational
  wave radiation. The expected rotation rates of the black holes are
  briefly discussed.
\end{abstract}


\section{INTRODUCTION}
\lSect{intro}

A large fraction of massive stars are found in binary systems with
sufficiently small separations that interaction is likely sometime in
the star's life \citep{Kim12,San11,San12}. This interaction will
  radically affect the sorts of supernovae they produce
  \citep{Pod92,Pol02,Wel99,Lan12,Dem17}. Many of the supernovae will no
  longer be Type II, but Type I. More subtle changes also happen to 
  the core structure that affect the nature of its explosion, including 
  energy, nucleosynthesis, and remnant properties.

Realistic studies of the evolution of stars in binaries can be
quite complicated.  In addition to the usual uncertainties in
mass-loss rates, mixing, rotation, and explosion physics inherent in
any study of massive stars, there is the added complexity of mass and
angular momentum exchange between the two components. This history
depends, in turn, on the initial orbital parameters, mass ratios,
efficiency of mass transfer, and the uncertain outcome of common
envelope evolution. Kicks may also be important in determining the
orbital parameters after each of the two explosions \citep{Van20}.

Using approximations for some of these uncertainties, many
  previous papers have estimated the properties of the neutron stars
  and black holes massive binaries leave behind
  \citep[e.g.][]{Bel02,Dem17,Fry12,Bel12,Dom12,Dem15,Eld16,Eld17,Bel20}. While
  exploring the effects of binary membership quite well, the treatment
  of the supernova explosion itself was often simplistic in these
  works. Presupernova models were sometimes adopted from several
  sources that used different physics to study the evolution. The
  range of masses considered was often limited and mass loss by winds
  was treated using prescriptions that, in some cases, have become
  dated. The role of different assumptions regarding the effects of
  binary membership was sometimes hard to disentangle.
  
Here, a different tack is taken \citep[see
    also][]{Woo19,Ert20}. The ``model'' for binary mass exchange is
  trivial and not intended as a substitute for more realistic
  calculations using population synthesis. It is assumed that the
  chief effect of evolution in a close mass-exchanging system is to
  remove the hydrogen envelope when the star first attempts to expand
  to red or blue supergiant proportions. This expansion is assumed to
  occur near the time of central helium ignition. As a consequence of
  losing its envelope, the helium core shrinks due to mass loss by a
  wind rather than growing as hydrogen burns in a surrounding
  shell. Thus the presupernova stars, in this case the residual cores
  of helium and heavy elements, are smaller in the ``binary'' case
  than for single stars. This result agrees, qualitatively, with the
  observation that Type Ib and Ic supernovae are more tightly
  correlated with star forming regions than Type IIp
  \citep{Kun18}. Hence they come from stars of greater main sequence
  mass. Quantitatively, the results of \citet{Woo19} are not so
  different, for a given main sequence mass, than those of
  \citet{Yoo10} who treated binary evolution more realistically. For
  example a 25 \Msun \ solar metallicity main sequence star leads to a
  presupernova core mass of 4 to 6 \Msun \ in Table 1 of
  \citet{Yoo10}. For \citet{Woo19}, the presupernova mass is 5.0
  \Msun. For an isolated 25 \Msun \ star that kept part of its
  hydrogen envelope until death, the helium core mass would have been
  8.4 \Msun\ \citep{Suk18}. Since the explosion properties and remnant
  masses depend sensitively on the presupernova core mass, the
  difference is substantial.

The ansatz of prompt envelope removal facilitates a calculation
  that otherwise would have been complex and highly parametrized, but
  also restricts the applicability of the results to just those stars
  initially close enough to quickly lose their envelopes. We do not treat
  here stars that would have been Type II supernovae of any sort. Any
  residual hydrogen is presumed to be promptly lost to a wind.
  Angular momentum transport is not followed and thus we cannot
  calculate the final spin of the core (though see
  \Sect{conclude}). The complications of uncertain mass loss in
  extremely massive stars that never become red supergiants, but instead
  make luminous blue variables are ignored. Every massive star will
  have a well defined helium core mass when it dies, but the relation
  between that final mass and main sequence mass is especially
  uncertain for such stars. We assume the mapping given by eqns. 3 and
  4 of \citet{Woo19}.  For very high masses the initial helium core
  mass is roughly half the main sequence mass.

Perhaps the worst error is ignoring the effect of mass exchange
  on the {\sl initial} helium core mass of the {\sl secondary}. The
  helium core mass of the secondary, at helium ignition, will be
  larger than if it had evolved in isolation since it will have
  accreted some uncertain fraction of its companion's envelope. The
  secondary may also transfer mass to the compact remnant that
  resulted from the death of the primary star. The histories of these
  accretion processes depend on the masses and metallicities of the
  stars, the orbital parameters, how much mass is lost from the system
  during any common envelope interaction, and the natal kick of the
  primary star's remnant.  These uncertainties and other
  characteristics of real binary systems are neglected here.  It is
  assumed that both stars produce presupernova cores defined only by
  their own initial mass and the adopted mass-loss rate. Consequently,
  we will underestimate the final masses of half of the
  stars. Crudely, the effect is that of changing the IMF for the 
  main sequence so as to favor the production of more massive stars, i.e., the
  effective slope will be be less steep than Salpeter.
  
With these caveats, what is really studied here is just the
  remnant mass distributions resulting from a library of mass-losing
  helium stars whose evolution and explosion were calculated in two
  previous papers \citep{Woo19,Ert20}. A semi-analytic approach to
  stellar evolution similar to \citet{Hur00} is used to estimate the
  presupernova mass distribution resulting from a given initial
  distribution of helium core masses at helium ignition. This
  semi-analytic description gives results in excellent agreement with
  the full stellar models calculated by \citet{Woo19} and can be used
  to estimate how those results would change for different
  metallicities and mass loss prescriptions without running actual
  stellar evolution calculations. It is further assumed that the
  remnant mass distribution is uniquely determined by the presupernova
  masses of the stripped cores, and that mapping is not sensitive
  to metallicity or interior composition. The mapping is determined
  from a grid of 1D neutrino-transport models calculated by
  \citet{Ert20}. Our study includes the full range of stellar masses
  expected to produce black holes and most of the stars that make
  neutron stars. The progenitors of electron--capture supernovae and
  the results of stellar mergers are omitted.

\section{Procedure}
\lSect{proc}

To begin, we derive the distribution of presupernova masses for a
given initial mass function (IMF), taken here to be Salpeter--like
\citep{Sal55}. That is done by integrating the mass-loss equation,
$dM/dt = f(M,L,Y,Z_{\rm init})$ where $M$ is the mass, $L$ the
luminosity, $Y$ the surface helium mass fraction, and $Z_{\rm init}$,
the initial metallicity of the star, not counting any newly
synthesized heavy elements that appear at the surface later due to
mass loss. $M$, $L$, and $Y$ vary with time, but not $Z_{\rm
    init}$. For mass loss it is the mass fraction of the iron group
  that matters most \citep{Vin05}. It is implicitly assumed here that
  the ratio of iron to other metals remains constant for the range of
  metallicities considered, chiefly \Zsun \ and 0.1 \Zsun. The initial
  stars are assumed to be composed of only helium plus $Z_{\rm
    init}$. For solar metallicity the abundances of \citet{Lod03} are
  used, for which the mass fraction of elements heavier than helium is
  0.0145 and for iron-group elements (mostly $^{56}$Fe) is 0.00147.

Three recent determinations of mass-loss rates for stripped
  helium stars are considered (\Fig{mdot}), taken from \citet{Yoo17},
\citet{Vin17}, and \citet{San19a}. The work of Yoon is a refinement of
earlier empirical fits while the two studies by Vink and Sander et al.
are of a ``first principles'' nature, based on calculations of
radiation transport and hydrodynamics. The former might naturally
resemble more observations in the mass range where it is fit while the
latter two might be more safely extrapolated to lower mass where the
work by Yoon lacks adequate calibration \citep{Gil19}.

Yoon gives two formulae for WN and WC stars. For WN stars, 
\begin{multline}
\log \, \dot M_{\rm Yoon, WN} \ = \ -11.32 \, + \, 1.18 \, \log
\left(\frac{L}{\Lsun}\right) \cr + 0.6 \, \log \, \left(\frac{Z_{\rm init}}{\Zsun}\right),
\lEq{mdotwne}
\end{multline}
which is taken from \citet{Nug00}. For WC stars, he gives
\begin{multline}
\log \, \dot M_{\rm Yoon, CO} \ = \  -9.2 \, + \, 0.85 \, \log
\left(\frac{L}{\Lsun} \right) \, + \cr 0.44 \, \log \, Y \,
  +  0.25 \, \log \, \left(\frac{Z_{\rm init}}{\Zsun}\right),
\lEq{mdotwr}
\end{multline}
which is taken from \citet{Tra16}. Yoon multiples both equations by a
factor, $f_{\rm WR}$, to account for uncertainty, and favors a value
$f_{\rm WR}$ = 1.58. Here we treat $f_{\rm WR}$ = 1 as the standard
case, but also explore $f_{\rm WR}$ = 2.

Note the explicit dependence in the second equation on $Y$. This does
not reflect the physical role of helium in absorbing radiation, but is
an empirical adjustment for the effect of increasing the carbon and
oxygen mass fractions.  Yoon's formula is strictly valid only for $Y <
0.9$, but, to good approximation, we can take it to characterize any
star that has lost its unburned helium shell.

The rate from \citet{Vin17} is for optically thin, helium-stripped
stars. Designed to reflect stars on the order of 4 \Msun, it is
possibly a severe underestimate for more massive helium cores at solar
metallicity. At lower metallicity however, the regime of lower mass
loss reflected by \citet{Vin17} could stretch to much higher
masses. Here it is used as a lower bound in all calculations,
especially for the \citet{San19a} rate which otherwise goes to very
small values at low mass.  Nominally for WN-type stars, Vink gives
\begin{equation}
\log \dot M_{\rm Vink} \ = -13.3 \ + \ 1.36 \, \log
\left(\frac{L}{\Lsun}\right) + 0.61 \log \left(\frac{Z_{\rm init}}{\Zsun}\right).
\end{equation}

\citet{San19a} give interpolation formulae as a function of $L/M$ for
WN and WC stars individually for two different metallicities in their
Table 3. See also their Fig. 20. In the general case, one would need
to interpolate for values of metallicity other than \Zsun \ and 0.1
\Zsun. Here we restrict our survey of Sander et al. rates to just
those two values of metallicity.  The three mass-loss rates used are
plotted as a function of the current luminosity of the helium star for
the two metallicities considered in \Fig{mdot}. The Vink and Yoon
rates are smooth, almost linear functions of the luminosity, while the
Sander et al rate shows a steep cut off at low luminosity owing to its
strong sensitivity to the Eddington parameter. As previously noted,
the \citet{Vin17} is taken as a lower bound to the \citet{San19a}
mass-loss rate for low luminosity.

% fig 1 - mass-loss rates
\begin{figure}
\includegraphics[width=0.48\textwidth]{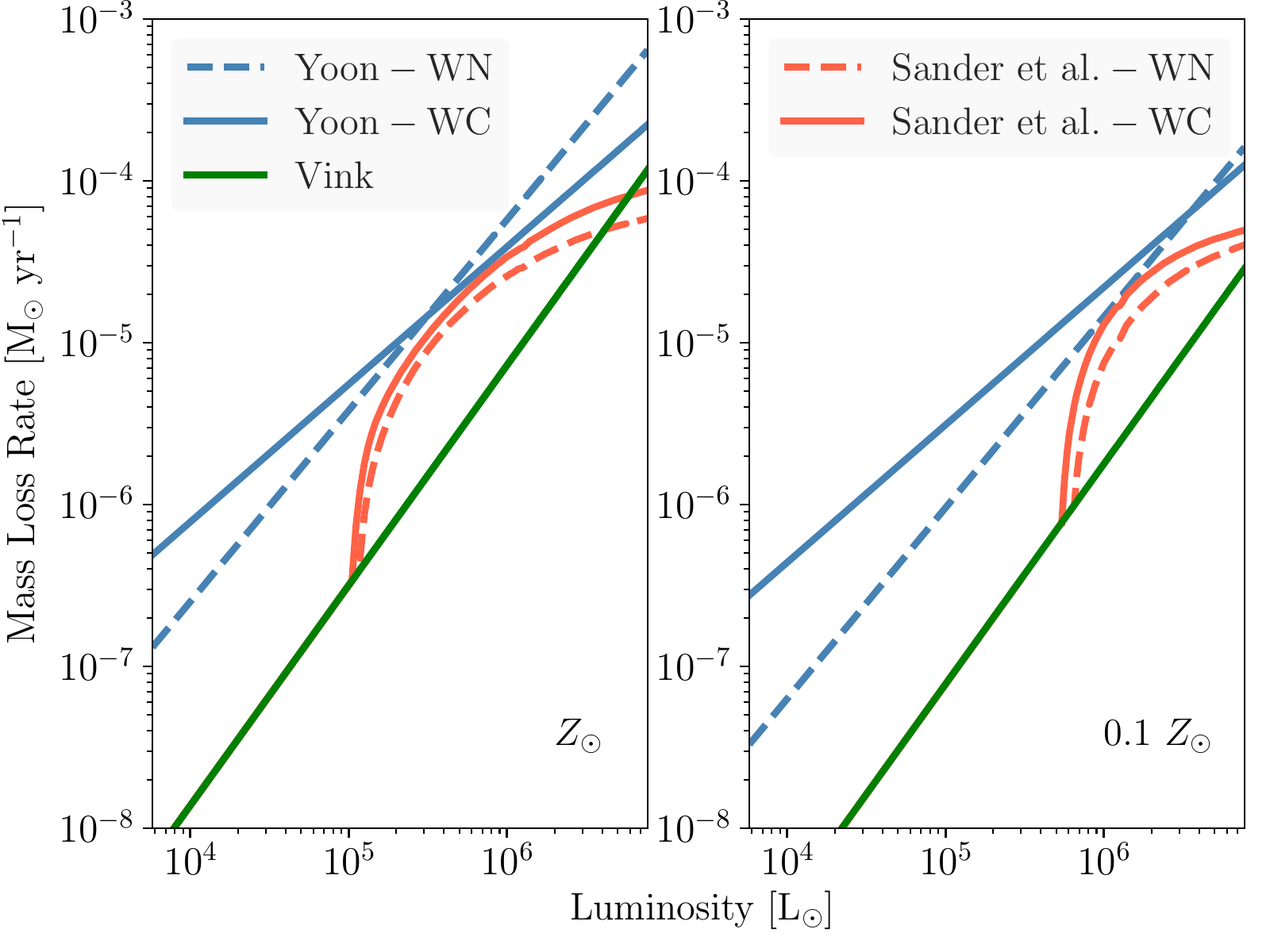}
\caption{Mass loss as a function of luminosity for WN and WC stars of
  solar and 10\% solar metallicity. The expressions plotted are from
  \citet[blue]{Yoo17}, \citet[green]{Vin17}, and \citet[red]{San19a}. 
  Blue and red dashed lines are for WN stars, while the Vink rate makes 
  no distinction between WN and WC stars. The rate from Vink has been 
  used, for low values of mass and luminosity, as a lower bound for 
  the rate from Sander et al. 
  \lFig{mdot}}
\end{figure}

To integrate these equations for a given metallicity, one must specify
the evolution of the luminosity and surface helium abundance. The
latter is also used as a switch to determine whether to use WN or WC
mass loss in the cases where two descriptions are given. For higher
mass helium stars, the extent of the helium burning convective core
grows. In the models of \citet{Woo19}, for low initial mass near 4
\Msun, the helium convective core in the presupernova models is only
about 52\% of the star's mass, but for helium core masses above 50
\Msun\ this fraction increases to near 85\% and stays relatively
constant. Between 4 and 50 \Msun, to reasonable approximation, the
fraction is obtained by interpolation,
\begin{equation}
\frac{M_{\rm WN-WC}}{M_{\rm He,i}} \ \approx \left(0.52 + 0.05
(M_{\rm He,i} -4)^{1/2}\right),
\lEq{WN-WC}
\end{equation}
where $M_{\rm He,i}$ is the initial mass of the helium star and
$M_{\rm WN-WC}$ is the maximum extent, in mass, of the convective
helium core. This mass also marks the transition point for using WC
instead of WN rates. Once the processed core is revealed, the surface
helium abundance in real stellar models declines rapidly to a minimum
value near 0.2 \citep[Table 4 of][]{Woo19}. The decline is not
precipitous though because the receding helium convective core leaves
behind a gradient that the mass loss eventually reveals. Adding
this gradient is not a large effect, but its inclusion mildly improves
the agreement with the models.  Variation of $Y$ from 0.8 to 0.2 only
changes the Yoon mass-loss rate by a factor of two and this variation
only occurs during a small fraction of the lifetime. Once the current
mass of the star, $M(t)$, fell below $M_{\rm WN-WC}$, we used the
approximation
\begin{equation}
Y \ \approx 0.2 \left(\frac{M(t)}{M_{\rm He,i}}\right)^2.
\end{equation} 

The luminosity history must also be approximated.  Fortunately, the
luminosity of a helium star during central helium burning and its
lifetime are mostly determined by its current mass and not very
sensitive to how much helium has burned. For example, a 10 \Msun
  \ helium star evolved without mass loss has a luminosity of
  10$^{5.15}$ \Lsun, 10$^{5.21}$ \Lsun, and 10$^{5.28}$ \Lsun\ after
  burning 2\%, 50\%, and 90\% respectively of the helium in its
  center. For a 4 \Msun \ helium star, the corresponding numbers are
  10$^{4.24}$, 10$^{4.32}$ and 10$^{4.34}$ \Lsun. Taking the
  luminosity from only the 50\% depletion model thus makes a maximum
  error in the instantaneous luminosity for most helium abundances of
  about $\sim$25\%. The average error is less. The helium burning lifetime
is also sensitive to the mass and not much to the metallicity. If, for
a given metallicity, the mass-loss rate and the lifetime are functions
of only the current mass, then the mass-loss equation can be
integrated numerically, for a given initial helium core mass to give
the final presupernova mass.

\begin{deluxetable}{rclrcl}
\tablecaption{Constant Mass Helium Stars}
\tablehead{ \colhead{${M_{\rm He}}$}  &
            \colhead{$\log\ (L)$}   &
            \colhead{\quad$\tau$}        &
            \colhead{${M_{\rm He}}$}  &
            \colhead{$\log\ (L)$}   &
            \colhead{\quad$\tau$}     
             \\
            \colhead{[\Msun]}  &
            \colhead{[\Lsun]}  &
            \colhead{[10$^{13}$ s]} &
            \colhead{[\Msun]}  &
            \colhead{[\Lsun]}  &
            \colhead{[10$^{13}$ s]} 
            }
\startdata
  2.0 &  3.47 &  14.2 &  12.5 &  5.39 &   1.85  \\
  2.1 &  3.54 &  12.9 &  13.0 &  5.42 &   1.81  \\
  2.2 &  3.60 &  11.9 &  13.5 &  5.45 &   1.77  \\
  2.3 &  3.66 &  11.0 &  14.0 &  5.48 &   1.74  \\
  2.4 &  3.71 &  10.3 &  16.0 &  5.58 &   1.62  \\
  2.5 &  3.76 &  9.65 &  18.0 &  5.66 &   1.53  \\
  2.8 &  3.88 &  8.00 &  20.0 &  5.73 &   1.45  \\
  3.0 &  3.99 &  7.00 &  22.0 &  5.80 &   1.39  \\
  3.5 &  4.17 &  5.95 &  24.0 &  5.86 &   1.33  \\
  4.0 &  4.32 &  5.04 &  26.0 &  5.91 &   1.30  \\
  4.5 &  4.45 &  4.39 &  28.0 &  5.96 &   1.26  \\
  5.0 &  4.57 &  3.91 &  30.0 &  6.00 &   1.23  \\
  5.5 &  4.66 &  3.54 &  32.0 &  6.04 &   1.20  \\
  6.0 &  4.75 &  3.26 &  34.0 &  6.08 &   1.18  \\
  6.5 &  4.83 &  3.03 &  36.0 &  6.10 &   1.16  \\
  7.0 &  4.90 &  2.84 &  38.0 &  6.14 &   1.14  \\
  7.5 &  4.96 &  2.67 &  40.0 &  6.17 &   1.12  \\
  8.0 &  5.02 &  2.54 &  50.0 &  6.30 &   1.05  \\
  8.5 &  5.07 &  2.42 &  60.0 &  6.41 &   1.00  \\
  9.0 &  5.12 &  2.32 &  66.0 &  6.46 &   0.98  \\
  9.5 &  5.17 &  2.23 &  70.0 &  6.49 &   0.97  \\
 10.0 &  5.21 &  2.15 &  80.0 &  6.57 &   0.94  \\
 10.5 &  5.25 &  2.08 &  90.0 &  6.63 &   0.92  \\
 11.0 &  5.29 &  2.01 & 100.0 &  6.68 &   0.91  \\
 11.5 &  5.33 &  1.96 & 120.0 &  6.78 &   0.88  \\
 12.0 &  5.36 &  1.91 & 150.0 &  6.89 &   0.86  
\enddata
\tablecomments{Helium stars are evolved at constant mass to build the 
  relation for $L(M)$. The luminosity is evaluated at 50\% helium 
  depletion. $\tau$ is the time from helium ignition until a central 
  temperature of $5 \times 10^8$ K is reached shortly before carbon 
  ignition.}
  \lTab{hetable}
\end{deluxetable}

To evaluate the luminosity as a function of mass and the lifetime, a
grid of 52 constant mass helium stars ranging from 2 to 150 \Msun\ was
evolved to carbon ignition using the implicit hydrodynamics code
\KEPLER\ \citep[similar setup as in][]{Woo19}.  The luminosity at 50\%
helium depletion and the total helium burning lifetime were tabulated
for each (\Tab{hetable}). The mass--dependent luminosity, $L(M)$, drives
the evolution of stellar mass, $dM/dt = f(L,M,Y,Z_{\rm init})$, which
was integrated, for a given metallicity and various mass-loss
prescriptions.  For greater accuracy, the remaining lifetime was
varied as the mass of the star decreased. This was accomplished by
increasing the time step in proportion to the current remaining lifetime.

Given the luminosity as a function of mass and the mass-loss rates as
a function of luminosity, mass, and metallicity, the mass-loss
equation was then integrated. On a laptop this took about a minute for
a grid of 30,000 masses between 2.5 and 150 \Msun. Four different
variations of mass-loss rate and two metallicities were
considered. \Fig{calibrate} shows a comparison of the semi--analytic
results and the presupernova masses derived for actual stellar models
using the \KEPLER\ code and Yoon mass-loss rate \citep{Woo19}. Errors
are small for low and moderate mass-loss rates, but increase slightly
for larger values due to the approximate treatment of the surface
helium abundance and luminosity in our equations. Both could be
improved.

% fig 2 - presupernova mass calibration
\begin{figure}
\includegraphics[width=0.48\textwidth]{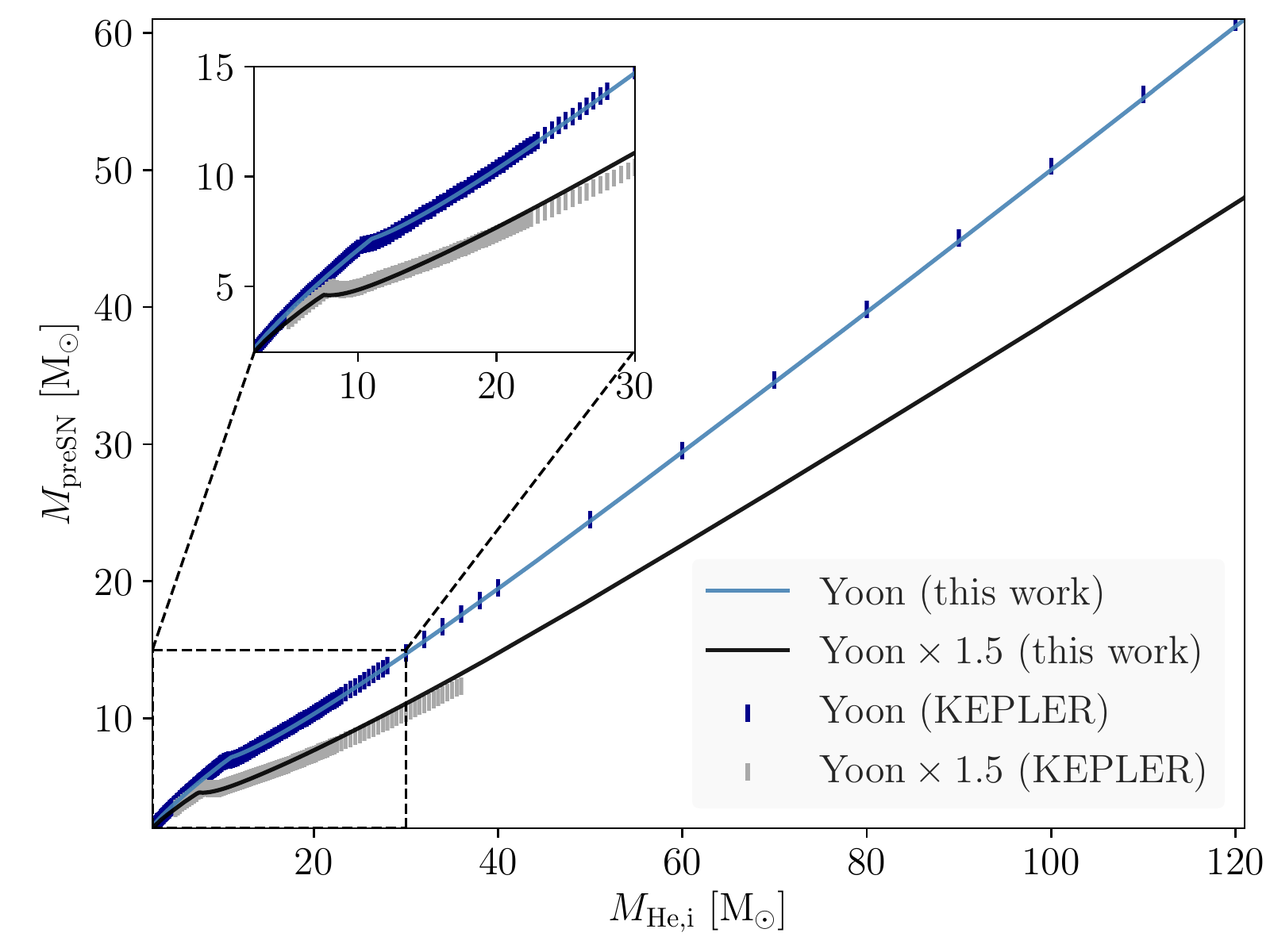}
\caption{Presupernova masses calculated using the analytic approach
  described in this paper (solid curve) compared with values for
  actual stellar models of the same initial mass, metallicity and 
  mass-loss rate, calculated using the \KEPLER\ code 
  \citep[vertical bars]{Woo19}. The standard Yoon mass-loss rate 
  (blue) and a value 1.5 times larger (gray) are considered. 
  Inflection points where the curves change slope at around 8 
  \Msun (gray) and 10 \Msun\ (blue) reflect the transition from WN 
  to WC stars.
  \lFig{calibrate}}
\end{figure}

% fig 3 - presupernova masses 
\begin{figure}
\includegraphics[width=0.48\textwidth]{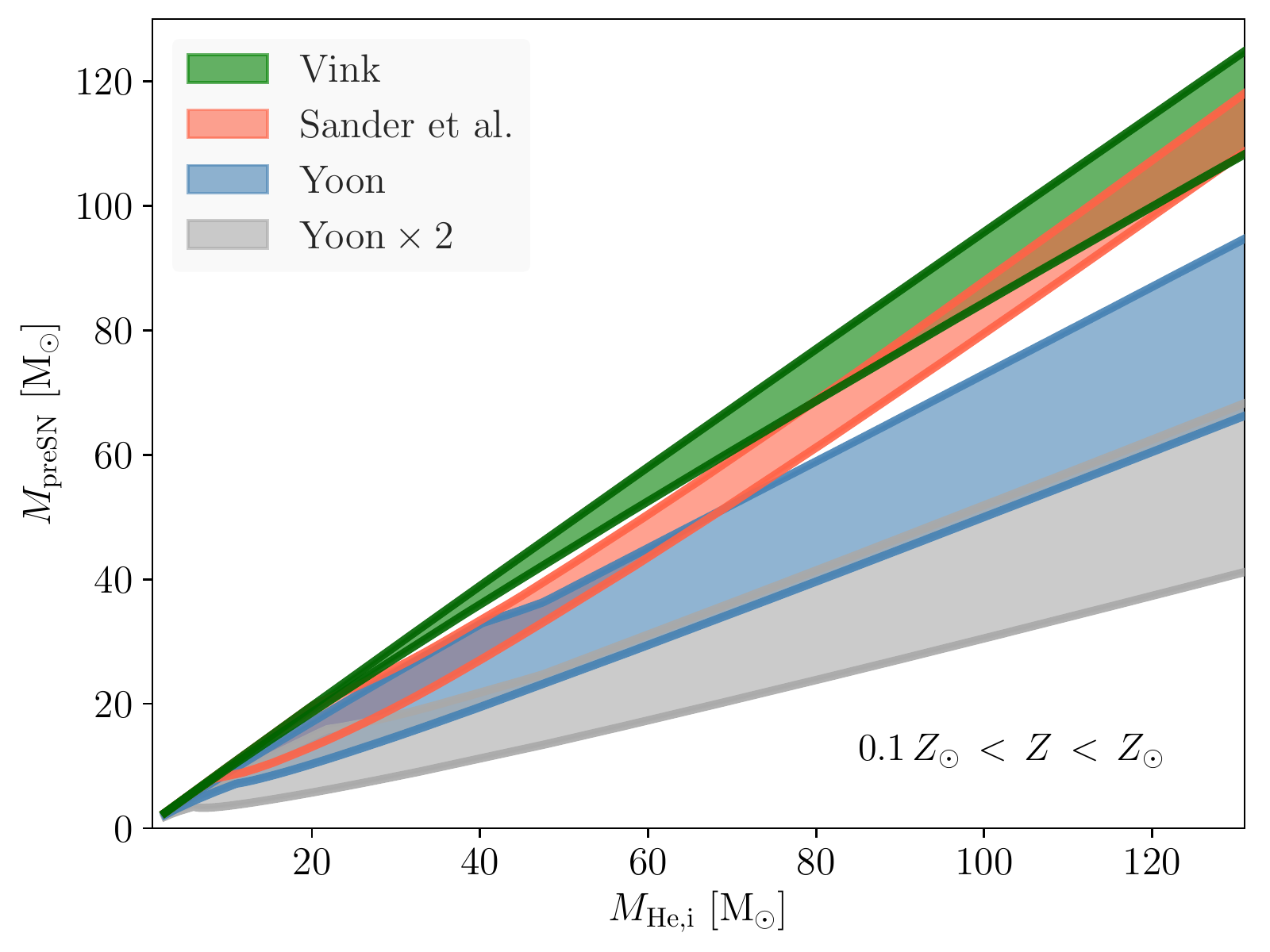}
\caption{Initial and final masses for the mass-loss rates shown 
  in \Fig{mdot}. Each color band is bound on the bottom by solar 
  metallicity and on the top by 10\% solar solar metallicity 
  stars. The mass loss is negligible for the Vink rate at 10\% 
  solar metallicity.
  \lFig{remsol}}
\end{figure}

The resulting presupernova masses are shown for different
metallicities and mass-loss rates in \Fig{remsol} for two values of
metallicity, solar and 10\% solar, and four choices of mass-loss
prescription.  Smaller values of mass loss were not calculated since
the lower limit on the mass loss from Vink already gives presupernova
masses essentially equal to the initial mass, i.e., the mass loss was
negligible. The Vink results at 10\% solar metallicity can thus be
used to approximate all smaller values of metallicity, including zero.

The second stage of our calculation required mapping these
final presupernova masses into the remnant masses produced when their
iron cores collapse. This mapping depends, of course, on the uncertain
explosion model. For stars that might have some reasonable chance at
exploding by the neutrino--driven mechanism, we used the ``death
matrix'' extracted, for the \texttt{W18} central engine, from the
recent work of \citet{Ert20}. Those calculations followed the outcomes
of the core--collapse, by calibrated neutrino--driven explosion
models, for a dense grid of presupernova masses from 2.1 to 20 \Msun,
which correspond to initial helium core masses from 2.5 to 40 \Msun.
The calculated gravitational masses include the mass decrement due to
neutrino losses determined self-consistently using P-HOTB \citep[see
  Table 3 and Fig. 16 of][]{Ert20}. Slightly larger values, typically
2\% in the gravitational mass, would be obtained using the correction
of \citet{Lat01}.

For higher masses, up to a presupernova mass of 30 \Msun, the
presupernova stars were assumed to collapse to black holes without an
explosion, but with 0.15 \Msun\ subtracted to account for neutrino
emission prior to the formation of a black hole. This is consistent
with the correction found by \citet{Ert20} for their highest mass
models that made black holes ``promptly'' (i.e., without
fallback). Above a presupernova mass of 30 \Msun, besides the
adjustment for neutrino mass loss, corrections were also made for mass
ejected by the pulsational pair instability based on the models from
\citet{Woo19}. For presupernova masses above 60 \Msun, his stars
exploded and left no remnants.

The resulting remnant masses are given as a function of presupernova
mass in \Tab{death} and \Fig{remnant}.  As in \citet{Ert20},
we adopt a maximum baryonic neutron star mass to be 2.75 \Msun, which
approximately corresponds to a gravitational mass of $\sim$2.3
\Msun. The maximum black hole mass is 46 \Msun, as in \citet{Woo19}.
Larger values, up to 56 \Msun \ could be accommodated for smaller values of the
$^{12}$C($\alpha,\gamma)^{16}$O reaction rate \citep[][Woosley, in
  prep.]{Far19}. The S-factor used for this critical reaction here was 175 keV
b at 300 keV, slightly higher than what was suggested in \citet{deB17}.

\begin{deluxetable}{cccccc}
\tablecaption{The ``Death Matrix''}
\tablehead{ \colhead{${M_{\rm preSN}}$}  &
            \colhead{${M_{\rm rem}}$}   &
            \colhead{${M_{\rm preSN}}$}  &
            \colhead{${M_{\rm rem}}$}    &
            \colhead{${M_{\rm preSN}}$}  &
            \colhead{${M_{\rm rem}}$}    
              \\
           \colhead{[\Msun]}  &
           \colhead{[\Msun]}  &
           \colhead{[\Msun]}  &
           \colhead{[\Msun]}  &
           \colhead{[\Msun]}  &
           \colhead{[\Msun]}  
            }\\
\startdata
 2.07 &  1.24 &  5.98 &  1.37 & 10.56 &  1.96 \\
 2.15 &  1.25 &  6.05 &  1.41 & 10.70 &  3.92 \\
 2.21 &  1.25 &  6.12 &  1.42 & 10.81 &  1.76 \\
 2.30 &  1.28 &  6.20 &  1.51 & 10.91 &  2.31 \\
 2.37 &  1.29 &  6.26 &  1.50 & 11.02 &  4.11 \\
 2.45 &  1.26 &  6.33 &  1.48 & 11.13 &  1.86 \\
 2.52 &  1.26 &  6.40 &  1.38 & 11.23 &  2.29 \\
 2.59 &  1.29 &  6.47 &  1.41 & 11.34 &  2.56 \\
 2.67 &  1.30 &  6.54 &  1.40 & 11.44 &  2.23 \\
 2.74 &  1.35 &  6.61 &  1.36 & 11.55 &  6.46 \\
 2.81 &  1.37 &  6.67 &  1.36 & 11.66 & 11.14 \\
 2.88 &  1.38 &  6.74 &  1.36 & 11.88 &  9.52 \\
 2.95 &  1.33 &  6.87 &  1.42 & 12.10 & 11.71 \\
 3.02 &  1.34 &  6.91 &  1.53 & 12.32 & 11.93 \\
 3.09 &  1.33 &  6.95 &  6.42 & 12.54 & 12.17 \\
 3.16 &  1.36 &  6.99 &  6.62 & 12.74 & 11.06 \\
 3.22 &  1.35 &  7.04 &  1.50 & 12.98 & 11.22 \\
 3.29 &  1.34 &  7.10 &  1.58 & 13.21 & 11.06 \\
 3.36 &  1.33 &  7.17 &  1.61 & 13.44 & 13.18 \\
 3.42 &  1.33 &  7.24 &  1.64 & 13.66 & 13.42 \\
 3.49 &  1.39 &  7.33 &  6.80 & 13.89 & 13.68 \\
 3.57 &  1.39 &  7.42 &  1.47 & 14.81 & 14.64 \\
 3.65 &  1.36 &  7.51 &  1.47 & 15.74 & 15.61 \\
 3.73 &  1.38 &  7.61 &  7.08 & 16.68 & 16.54 \\
 3.81 &  1.39 &  7.71 &  1.50 & 17.63 & 17.49 \\
 3.89 &  1.41 &  7.81 &  7.43 & 18.59 & 18.45 \\
 3.98 &  1.43 &  7.90 &  4.88 & 19.56 & 19.41 \\
 4.05 &  1.32 &  8.00 &  7.68 & 20.53 & 20.38 \\
 4.13 &  1.34 &  8.11 &  7.84 & 21.51 & 21.36 \\
 4.21 &  1.56 &  8.21 &  8.00 & 22.50 & 22.35 \\
 4.29 &  1.51 &  8.31 &  8.07 & 23.49 & 23.34 \\
 4.37 &  1.38 &  8.41 &  8.20 & 24.48 & 24.33 \\
 4.44 &  1.42 &  8.47 &  8.25 & 25.49 & 25.34 \\
 4.52 &  1.38 &  8.59 &  8.34 & 26.50 & 26.35 \\
 4.59 &  1.39 &  8.70 &  8.41 & 27.51 & 27.36 \\
 4.67 &  1.41 &  8.80 &  8.50 & 28.52 & 28.37 \\
 4.75 &  1.40 &  8.87 &  8.55 & 29.53 & 29.38 \\
 4.82 &  1.41 &  8.99 &  8.63 & 30.56 & 30.36 \\
 4.90 &  1.64 &  9.09 &  8.72 & 31.57 & 31.35 \\
 4.97 &  1.46 &  9.18 &  8.82 & 32.60 & 32.35 \\
 5.04 &  1.49 &  9.28 &  8.91 & 33.63 & 33.30 \\
 5.12 &  1.36 &  9.38 &  9.01 & 34.66 & 34.08 \\
 5.19 &  1.41 &  9.45 &  9.08 & 36.83 & 36.02 \\
 5.26 &  1.47 &  9.58 &  9.22 & 39.38 & 37.28 \\
 5.34 &  1.49 &  9.67 &  1.41 & 41.95 & 38.02 \\
 5.41 &  1.48 &  9.78 &  1.41 & 44.54 & 38.61 \\
 5.48 &  1.50 &  9.88 &  1.43 & 47.13 & 40.61 \\
 5.56 &  1.50 &  9.98 &  1.45 & 50.22 & 42.78 \\
 5.63 &  1.46 & 10.07 &  1.52 & 52.82 & 45.84 \\
 5.70 &  1.38 & 10.17 &  1.51 & 55.43 & 44.65 \\
 5.77 &  1.44 & 10.29 &  1.52 & 57.42 & 41.20 \\
 5.84 &  1.48 & 10.39 &  1.78 & 60.12 &  3.51 \\
 5.92 &  1.43 & 10.48 &  1.82 &       &       \\
 \enddata
 \tablecomments{All remnant masses are gravitational. Presupernova
   masses below 2.07 \Msun \ were assumed to produce a 1.24 \Msun
   \ neutron star. The maximum neutron star mass is taken as 2.3 \Msun, 
   and presupernova stars above 60.12 \Msun \ left no remnant. The 
   table is largely based on W18 engine results of \citet{Ert20}.}
 \lTab{death}
\end{deluxetable}

% fig 4 - Remnant mass as a function of pesupernova mass
\begin{figure}
\includegraphics[width=0.48\textwidth]{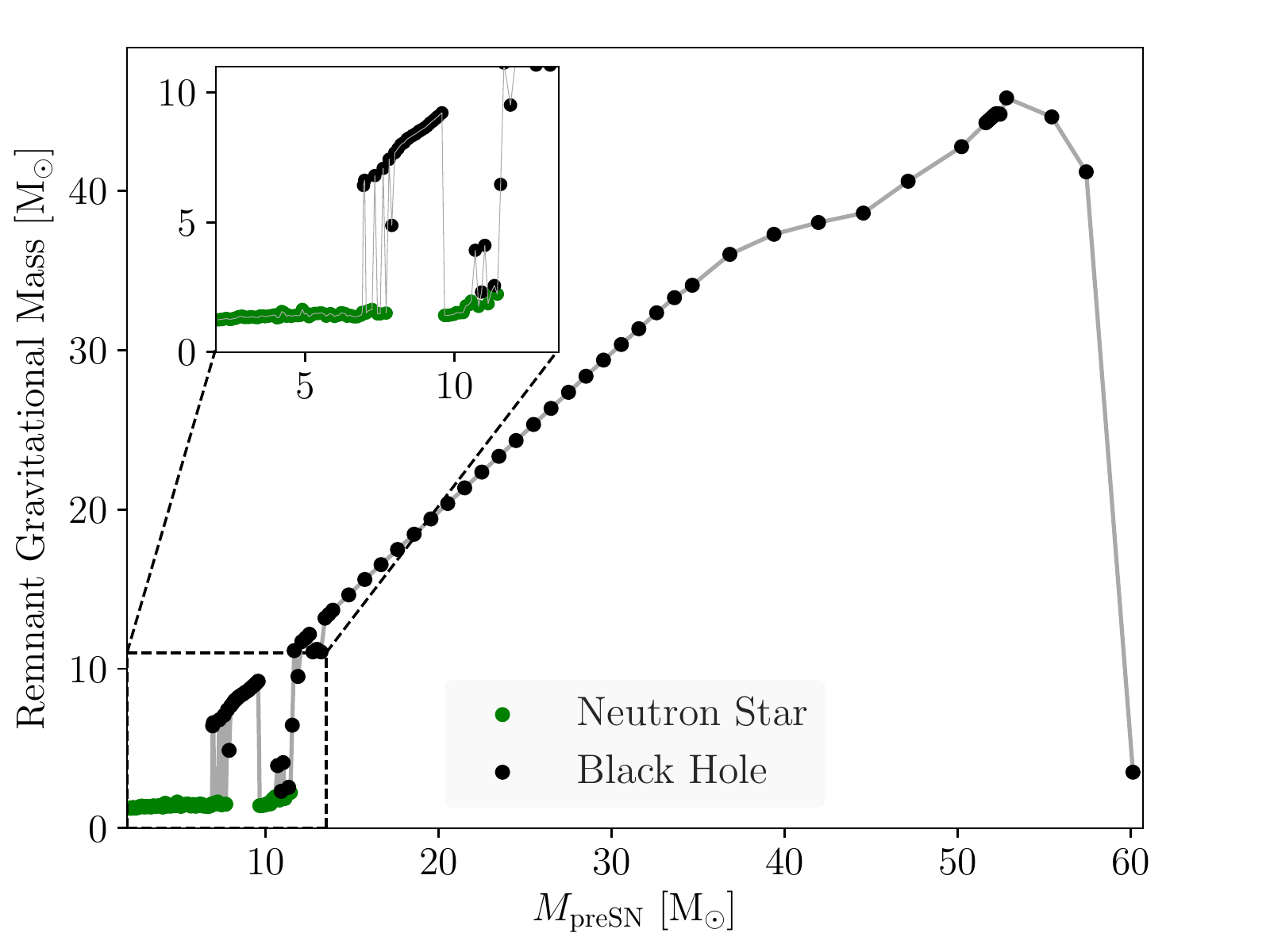}
\caption{Gravitational masses of neutron star (green) and black hole 
  (black) remnants as a function of presupernova mass based on the 
  results from \citet[W18 engine]{Ert20}. The presupernova stars were 
  assumed to have solar metallicity and evolved, including mass loss 
  \citep{Yoo17}, to the presupernova mass shown. The remnant masses 
  are corrected for neutrino mass loss \citep{Ert20}. The same 
  correspondence between presupernova mass and remnant mass is 
  assumed to exist for all choices of metallicity and mass-loss 
  prescription.
  \lFig{remnant}}
\end{figure}

\section{IMF Averaged Birth Functions for Compact Objects}
\lSect{imfaver}

The final stage consisted of weighting the remnant masses with an IMF
corresponding to their original main sequence progenitors. As in
\citet{Woo19}, it was assumed that, for initial main sequence masses of less 
than 30 \Msun, the helium core mass at the beginning of helium burning 
is
\begin{equation}
M_{\rm He,i} \ \approx \ 0.0385 \, M_{\rm ZAMS}^{1.603}  \, \Msun.
\lEq{mzams}
\end{equation}
For heavier stars
\begin{equation}
M_{\rm He,i} \ \approx \ 0.50 \, M_{\rm ZAMS} \, - 5.87 \, \Msun.
\lEq{mzamsh}
\end{equation}
The relation between initial helium star mass and presupernova mass is
given in \Fig{remsol}.

For a given semi--analytically computed presupernova mass, the 
corresponding remnant gravitational mass is obtained by interpolating 
on the relation shown in \Tab{death} and \Fig{remnant}. Any successful
supernova explosion that produced a neutron star or a black hole with 
a fallback mass greater than $10^{-2}$ \Msun\ was taken as a fallback 
case \citep[for details see][]{Ert20}. In order to avoid artificial 
structures in the birth function, we do \emph{not} interpolate between 
fallback and non--fallback cases, i.e. only interpolate if the nearest 
two grid points in \Tab{death} are either both fallback, or both 
non-fallback.  

\subsection{Neutron stars}
\lSect{nstar}

Combining these results, the birth function and its properties for
neutron stars are illustrated in \Fig{ns} and \Tab{nstbl}. A
Salpeter--like IMF with $\alpha=2.35$ was assumed over the entire mass
range (\Sect{proc}).  The distribution for solar metallicity and
mass-loss rates from \citet{Yoo17} agrees very well with the results
based on actual stellar models in \citet{Ert20}. The median
gravitational mass for $f_{\rm WR}$ = 1, i.e., 1.35 \Msun, is also
within 0.01 \Msun\ of their result. Perhaps this is not surprising
given that most of \Tab{death} was extracted from that work, but it
does validate the calculation of presupernova masses by simply
numerically integrating the mass-loss history.

\Fig{ns} and \Tab{nstbl} additionally show, however, a weak 
dependence of the distribution function on mass loss. For instance, 
the frequency of the most massive neutron stars ($>1.7$ \Msun), 
made by fallback in more massive progenitors \citep{Ert20}, is 
reduced with stronger mass-loss rate. For the low mass-loss rate 
from \citet{Vin17} the median neutron star mass is 1.37 \Msun\ and 
for the high mass-loss rate, twice that of the standard value of 
\citet{Yoo17}, the mass is 1.32 \Msun. We thus predict that the 
median gravitational mass of neutron stars in binary systems will 
vary, but only a little, with mass loss -- and hence with metallicity.

This agreement belies the fact that the {\sl total number} of neutron
stars is substantially different in the three cases. In each case, the
distribution function has been normalized so that the integral of
number over mass is unity. For the upper limit mass-loss rate (Yoon
with $f_{\rm WR}$ = 2), the fraction of core collapses that produce a
neutron star is 90\%, while for the lower limit mass-loss rate
(Vink's) it is only 69\% since many more stars remain massive enough
to collapse into black holes.

\Fig{ns} also shows the comparison with the standard case of Yoon
  mass loss rates with $f_{\rm WR} = 1$ and observationally inferred 
  distributions by \citet{Oze12} and \citet{Ant16}. For neutron star 
  gravitational masses below 1.5 \Msun, which are the bulk of the 
  observed objects, our model agrees well with both works though better 
  with \citet{Oze12}. We also agree with \citet{Ant16} on the existence 
  of a heavier component above 1.5 \Msun\ that, in our case, results 
  from fallback during the explosion. However, we produce far fewer 
  massive objects than claimed by \citet{Ant16}. Possible explanations are an 
  underestimate of fallback in our models or additional accretion by 
  the neutron star during the spin-up phase. This heavier component is 
  derived by \citet{Ant16} from observations of millisecond pulsars.

% fig 5 - neutron star masses for  Z solar
\begin{figure}
  \includegraphics[width=0.48\textwidth]{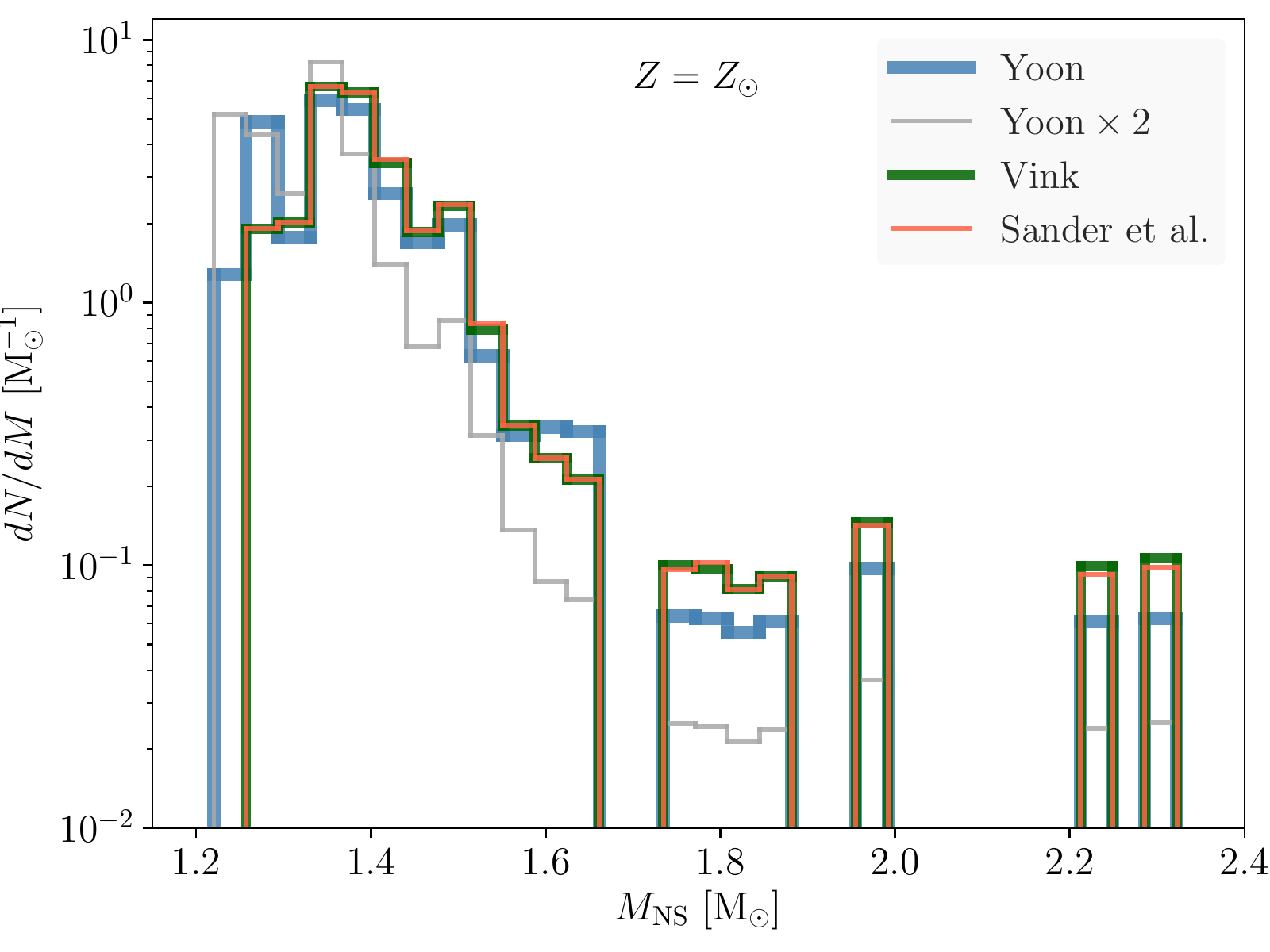}
  \includegraphics[width=0.48\textwidth]{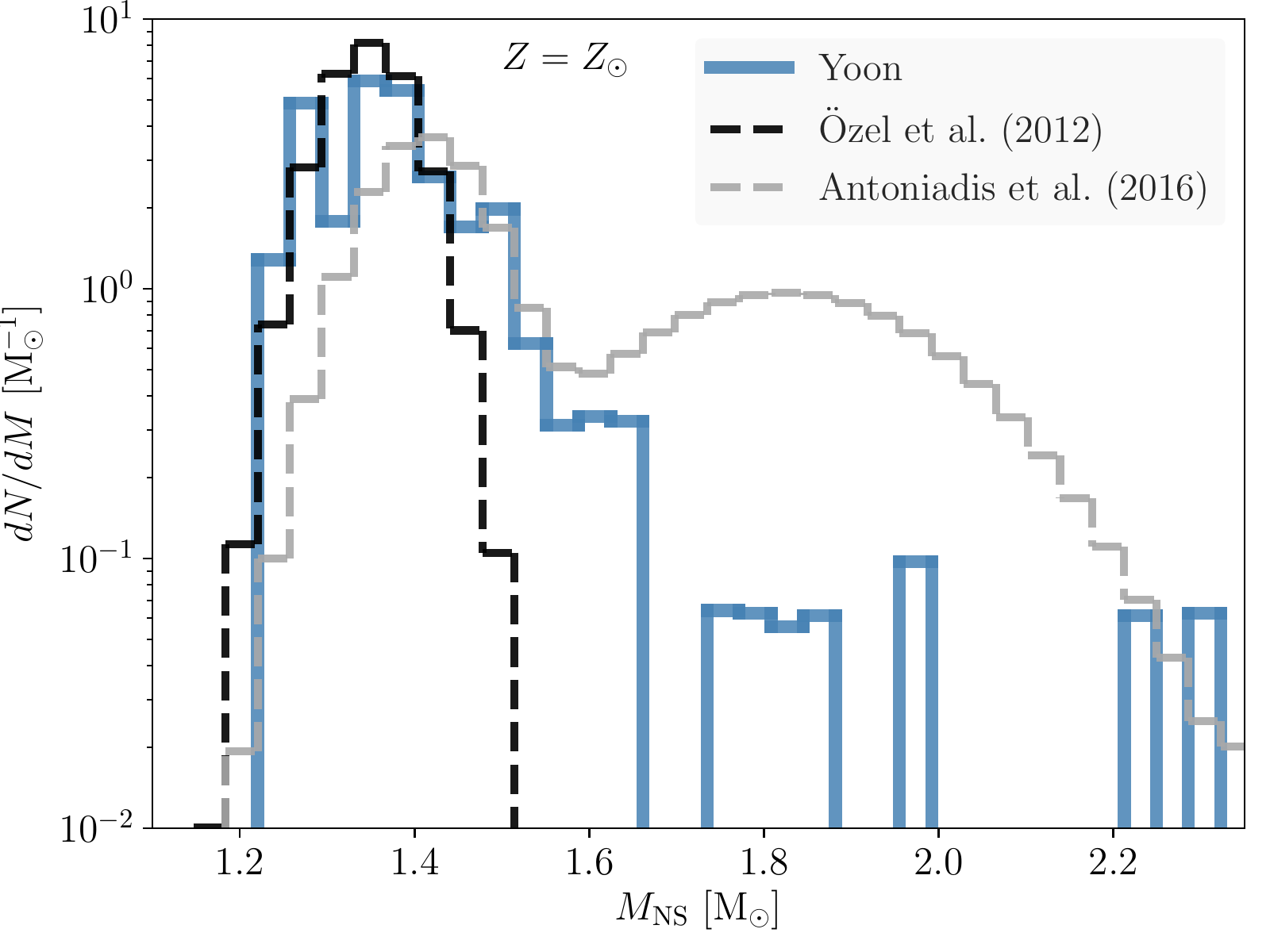}
\caption{Given the presupernova masses (\Fig{remsol}) for helium stars
  with solar metallicity evolved with mass loss (\Fig{mdot}), the
  remnant masses can be calculated from the data in \Fig{remnant} and
  \Tab{death}. The normalized birth function for the gravitational
  neutron star mass is shown here in the top panel for four choices of
  mass-loss rate, \citet[blue]{Yoo17}, \citet[green]{Vin17},
  \citet[red]{San19a}, and twice \citet[gray]{Yoo17}. The birth
  function is quite robust against changes in the mass-loss
  rate. Median values range from 1.32 to 1.37 \Msun, while the
  fraction of supernovae range from 90\% to 69\% between twice the
  Yoon's and Vink's prescriptions respectively. The bottom panel
  compares the case of Yoon mass loss with $f_{\rm WR}$ = 1 with
  observations by \citet{Oze12} and \citet{Ant16}. \lFig{ns}}
\end{figure}

\begin{deluxetable}{lcccl}
\tablecaption{Average Neutron Star Masses}
\tablehead{
            \multirow{2}{*}{$\dot{M}$}
          & \colhead{$M_{0.45}$}
          & \colhead{$\rm median$}
          & \colhead{$M_{0.55}$}
          & \colhead{$f_{\rm NS}$}\vspace{1mm}
          \\
          & \colhead{[\Msun]}
          & \colhead{[\Msun]}
          & \colhead{[\Msun]}
          & \colhead{}
          }
\startdata
& \multicolumn{4}{c}{$Z_{\rm init}=\Zsun$}\\
\vspace{-2mm}\\
\cline{2-5}\\
${\rm Yoon}$            & 1.341 & 1.349 & 1.359 & 0.784 \\
${\rm Yoon\times 2}$    & 1.314 & 1.320 & 1.326 & 0.907 \\
${\rm Vink}$            & 1.360 & 1.368 & 1.376 & 0.686 \\
${\rm Sander\ et\ al.}$ & 1.360 & 1.368 & 1.376 & 0.684 \\
\\
& \multicolumn{4}{c}{$Z_{\rm init}=0.1\ \Zsun$}\\
\vspace{-2mm}\\
\cline{2-5}\\
${\rm Yoon}$            & 1.355 & 1.364 & 1.372 & 0.700 \\
${\rm Yoon\times 2}$    & 1.349 & 1.357 & 1.366 & 0.715 \\
${\rm Vink}$            & 1.361 & 1.369 & 1.378 & 0.683 \\
${\rm Sander\ et\ al.}$ & 1.361 & 1.369 & 1.377 & 0.677
\enddata
\tablecomments{All quantities are evaluated with Salpeter
  $\alpha=2.35$ across the entire helium star mass range. 
  $M_{0.45}$ and $M_{0.55}$ are mass points where the 
  normalized fraction of neutron stars is 45\% below and 
  above respectively. $f_{\rm NS}$ is the fraction of 
  supernova explosions that form neutron stars.}
\lTab{nstbl}
\end{deluxetable}

\subsection{Black Holes}
\lSect{bh}

In a similar fashion, the birth function for black holes can be
calculated and is also found not to vary greatly with metallicity or
mass-loss prescription. A caveat is that it has been assumed that
helium cores all the way up to 150 \Msun, hence ZAMS stars to 300
\Msun \ are produced with the same continuous Salpeter--like mass
function. If star formation is truncated at a smaller mass, this would
appear as a cut off in the black hole mass distribution. If, for
example, no stars formed in close binaries with main sequence masses
over 100 \Msun, there would be no helium cores with initial mass over
44 \Msun. Hence for solar metallicity and standard Yoon mass-loss
rates, there would be no presupernova masses over 22 \Msun\ and no black
holes heavier than that either. Mass loss for extremely massive stars
at a rate very different from the unverified extrapolation of the
expressions used here, would also give different, interesting results.

\begin{deluxetable}{lcccl}
\tablecaption{Average Black Hole Masses}
\tablehead{
            \multirow{2}{*}{$\dot{M}$}
          & \colhead{$M_{0.45}$}
          & \colhead{$\rm median$}
          & \colhead{$M_{0.55}$}
          & \colhead{$f_{\rm BH}$}\vspace{1mm}
          \\
          & \colhead{[\Msun]}
          & \colhead{[\Msun]}
          & \colhead{[\Msun]}
          & \colhead{}
          }
\startdata
\vspace{-2mm}\\
\multicolumn{5}{c}{$2.5<M_{\rm He,i}<40$ [\Msun]}\\
\vspace{-2mm}\\
& \multicolumn{4}{c}{$Z_{\rm init}=\Zsun\quad\alpha=2.35$}\\
\cline{2-5}\\
${\rm Yoon}$            & 8.4  & 8.6  & 8.9  & 0.17 \\
${\rm Yoon\times 2}$    & 7.8  & 7.9  & 8.0  & 0.04 \\
${\rm Vink}$            & 13.2 & 14.2 & 15.2 & 0.30 \\
${\rm Sander\ et\ al.}$ & 8.4  & 8.6  & 8.9  & 0.29 \\
\\
& \multicolumn{4}{c}{$Z_{\rm init}=0.1\ \Zsun\quad\alpha=2.35$}\\
\cline{2-5}\\
${\rm Yoon}$            & 12.1 & 13.7 & 14.6 & 0.27\\
${\rm Yoon\times 2}$    & 10.9 & 11.8 & 13.3 & 0.24\\
${\rm Vink}$            & 13.6 & 14.5 & 15.6 & 0.30\\
${\rm Sander\ et\ al.}$ & 13.6 & 14.5 & 15.6 & 0.30\\
\cline{1-5} %%%%%%%%%%%%%%%%%%%%%%%%%%%%%%%%%%%%%%%%
\vspace{-1mm}\\
\multicolumn{5}{c}{$2.5<M_{\rm He,i}<150$ [\Msun]}\\
\vspace{-2mm}\\
& \multicolumn{4}{c}{$Z_{\rm init}=\Zsun\quad\alpha=2.35$}\\
\cline{2-5}\\
${\rm Yoon}$            & 9.5  & 11.3 & 13.1 & 0.22 \\
${\rm Yoon\times 2}$    & 11.5 & 13.2 & 14.3 & 0.09 \\
${\rm Vink}$            & 14.3 & 15.5 & 16.8 & 0.31 \\
${\rm Sander\ et\ al.}$ & 8.7  & 9.1  & 10.9 & 0.32 \\
\\
& \multicolumn{4}{c}{$Z_{\rm init}=0.1\ \Zsun\quad\alpha=2.35$}\\
\cline{2-5}\\
${\rm Yoon}$            & 14.1 & 15.3 & 16.5 & 0.30\\
${\rm Yoon\times 2}$    & 13.5 & 14.5 & 15.6 & 0.29\\
${\rm Vink}$            & 14.5 & 15.7 & 17.0 & 0.32\\
${\rm Sander\ et\ al.}$ & 14.7 & 16.0 & 17.3 & 0.32\\
\\
& \multicolumn{4}{c}{$Z_{\rm init}=\Zsun$}\\
\cline{2-5}\\
${\rm Yoon\ (\alpha=1.35)}$ & 16.4 & 18.1 & 20.0 & 0.47\\
${\rm Yoon\ (\alpha=1.85)}$ & 13.3 & 14.5 & 15.9 & 0.33\\
${\rm Yoon\ (\alpha=2.35)}$ & 9.5  & 11.3 & 13.1 & 0.22\\
${\rm Yoon\ (\alpha=2.85)}$ & 8.5  & 8.9  & 10.5 & 0.14\\
${\rm Yoon\ (\alpha=3.35)}$ & 8.2  & 8.4  & 8.6  & 0.08
\enddata
\tablecomments{All masses include correction to neutrino emission.
         $M_{0.45}$ and $M_{0.55}$ are mass points where the 
         normalized fraction of black holes is 45\% below and 
         above respectively. $f_{\rm BH}$ is the combined fraction 
         of implosions and supernovae that form black holes.}
         \lTab{bhtbl}
\end{deluxetable}

% fig 6 - BH masses solar Z
\begin{figure}
\includegraphics[width=0.48\textwidth]{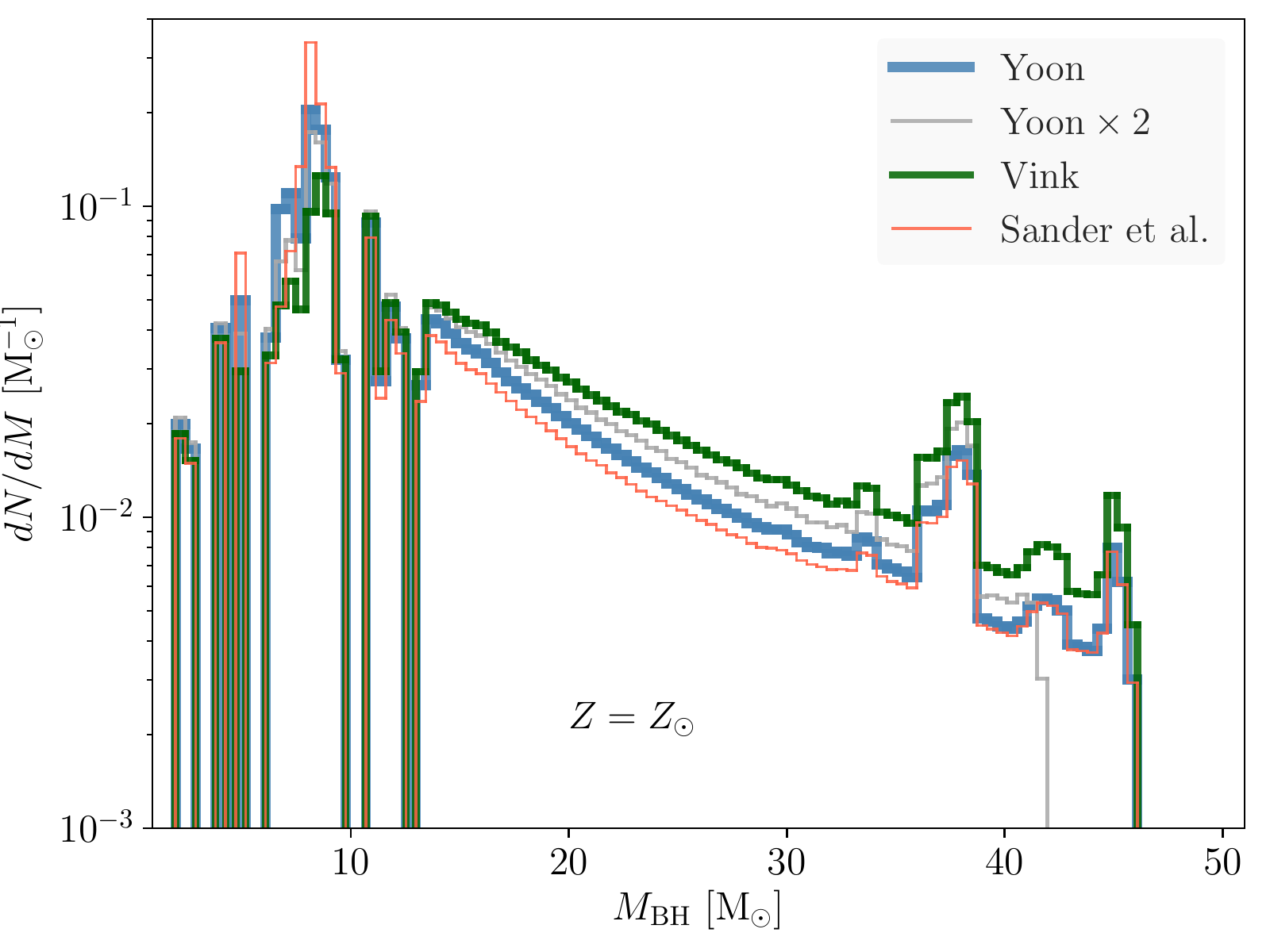}
\includegraphics[width=0.48\textwidth]{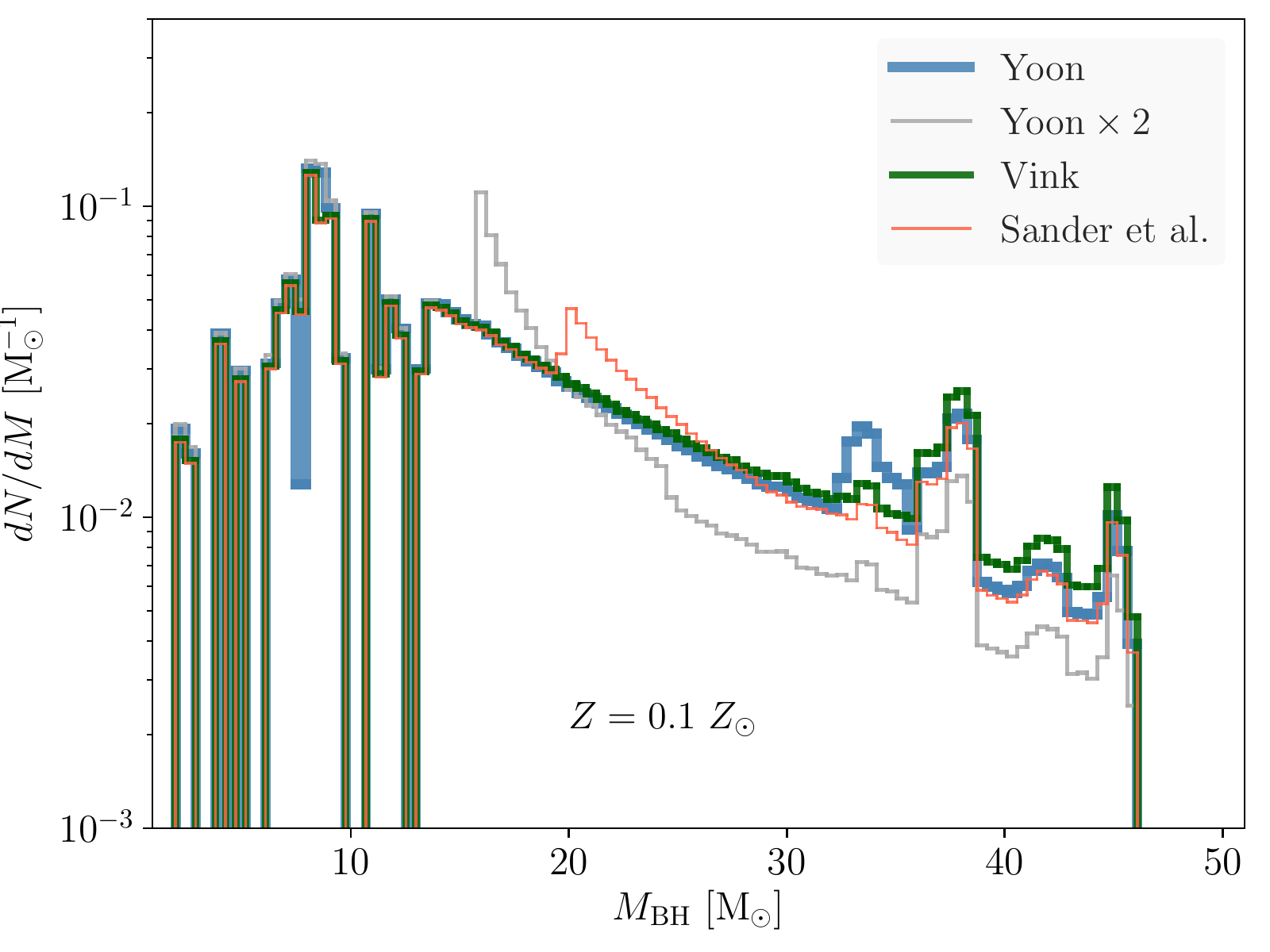}
\caption{The black birth function for solar metallicity (top) and 10\%
  solar metallicity (bottom) stars using the presupernova masses given
  in \Fig{remsol}.  The birth function is not very sensitive to the
  mass-loss prescription and metallicity. It is bounded on the lower
  end by the maximum mass neutron star and on the upper end by the
  onset of PPISN. There is no unpopulated ``gap'' between neutron
  stars and black holes. The gap near 10 \Msun\ is due to the
  presupernova compactness variation, and the pile--up at high mass is
  due to PPISN. At low metallicity, the Sander et al. prescription
  predicts a well defined pile--up near 20 \Msun. The peak of
    the distribution is broadly consistent with that inferred from
    X-ray binary systems \citep[e.g.,][]{Oze10,Far11}, and the range
    at high mass encompasses all currently known BH merger events from
    LIGO/Virgo within 90\% confidence level uncertainty.
  \lFig{bh}}
\end{figure}

The remnant mass distribution for black holes is shown in \Fig{bh},
and its properties are listed in \Tab{bhtbl} assuming that the mass
range of initial helium cores extends from 2.5 to 150 \Msun\ or 2.5 to
40 \Msun. The lower value corresponds to a still quite massive main
sequence star and is the limit of the \citet{Ert20} survey.
Besides the medians, which are not very precisely determined
due to the irregular shape of the distribution, the limits $M_{0.45}$
and $M_{0.55}$ are also given. These are the masses above and below
which 45\% of the black holes have their masses.  Helium star ranges
from 2.5 to 40 \Msun \ and 2.5 to 150 \Msun \ correspond to upper
limits of the main sequence mass function around 100 and 300 \Msun
\ respectively. For the more limited mass range, the median black hole
mass using standard Yoon mass loss ($f_{\rm WR}$ = 1) at solar
metallicity and the W18 central engine, which was used to make
\Tab{death}, is 8.6 \Msun. This is in excellent agreement with the 8.6
\Msun\ calculated by \citet{Ert20} using the same assumptions, but
actual stellar models, serving again to validate the
semi--analytic approach used here, i.e. the use of \Tab{hetable} to
integrate the mass-loss equation to estimate presupernova masses
without running actual stellar models. For the larger mass range,
including helium stars up to 150 \Msun, the results differ, though not
greatly. \citet{Ert20} gave a median black hole mass of 10.9
\Msun\ for the W18 engine. We get 11.3 \Msun. The new value is
slightly larger and more accurate because pulsational
pair--instability supernovae (PPISN) were not included in the previous
study.

The shape of the birth function (\Fig{bh}) is not very sensitive to
either metallicity or mass-loss rate. Note that all distributions have
been normalized so that the integral of $dN/dM$ over all black hole
masses is one. As in \citet{Ert20}, a few low mass black holes are
made by fallback and there is a narrow gap of production around 10
\Msun.  The exact location of this gap will be sensitive to the
compactness distribution of the presupernova stars and thus may vary
with different choices for convection theory and
$^{12}$C($\alpha,\gamma)^{16}$O reaction rate, but its presence is
robust and worth looking for, as it would provide constraints on the
presupernova evolution. The birth function for black holes left by
PPISN, i.e., masses 35 to 46 \Msun, is noisy. Whether this reflects
the sparse sample of PPISN used to generate the plot (Table 5 of
\citet{Woo19} has a resolution in black hole mass of about 2 \Msun) or
a physical effect needs further investigation.  Current results
suggest a preponderance of black holes with masses 35 to 40
\Msun. This is because the more massive PPISN near the upper limit (46
\Msun\ here) have more violent instabilities and eject more mass.

% fig 7 - BH masses Z dependence
\begin{figure}
\includegraphics[width=0.48\textwidth]{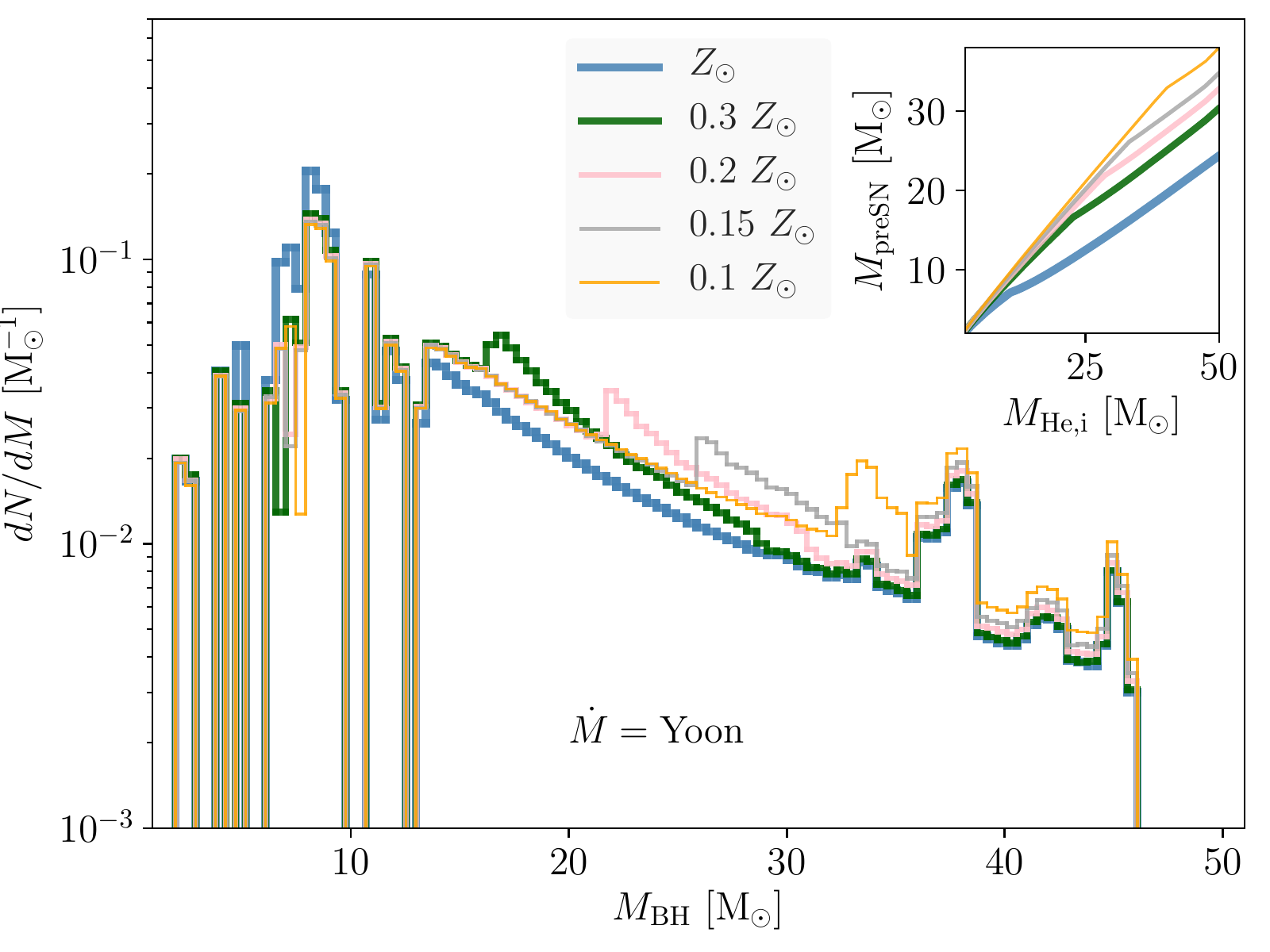}
\includegraphics[width=0.48\textwidth]{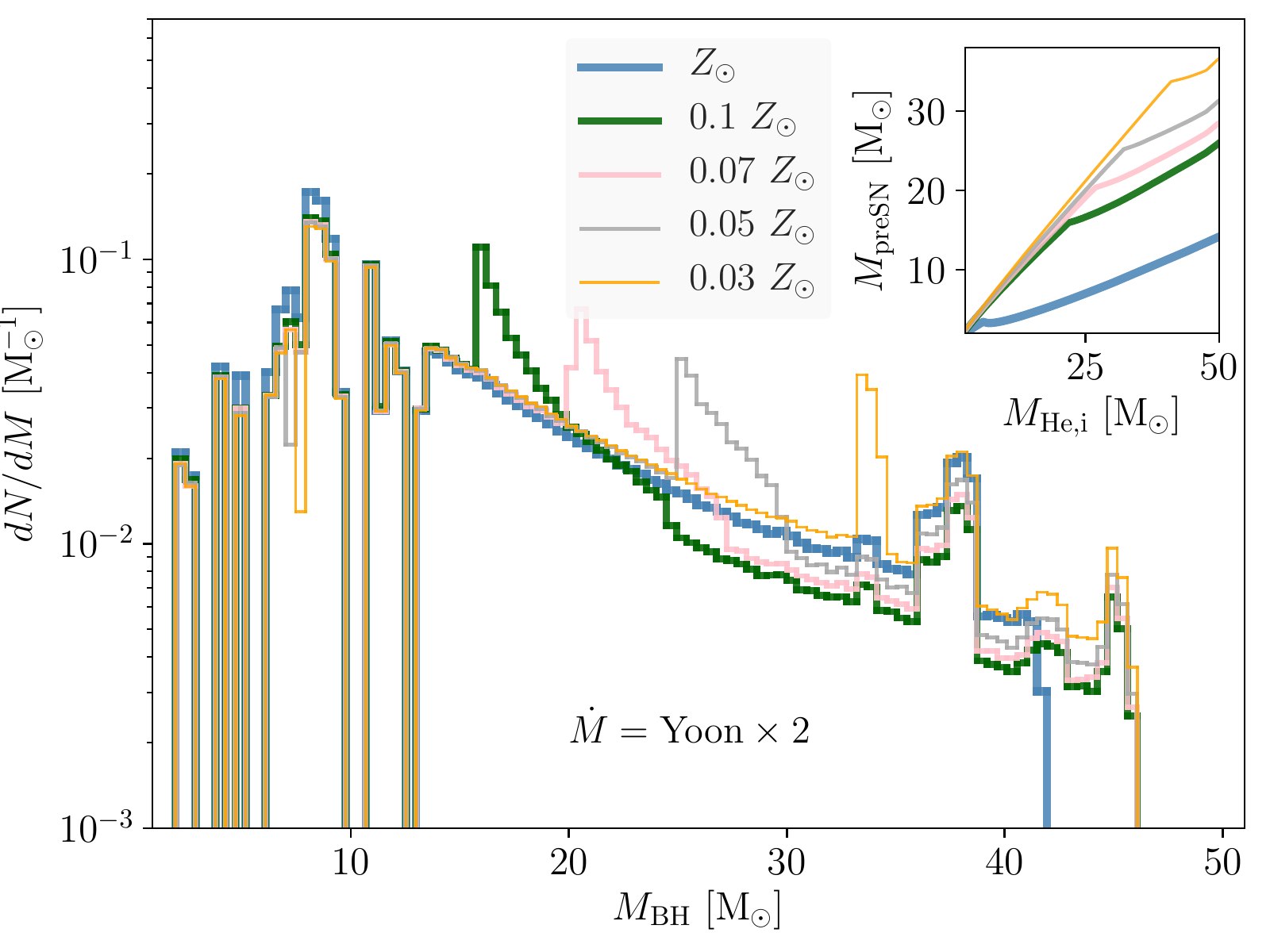}
\caption{Dependence of the black hole birth function on the
    metallicity for the Yoon mass loss rates with $f_{WR}$ = 1 (top
    panel) and 2 (bottom panel). Globally the shape of the birth
    function does not depend sensitively on metallicity though the
    number of black holes does (\Tab{bhtbl}). Note that each
    distribution is normalized so that the area beneath is one.  The
    sharp peak seen for some cases between 15 and 35 \Msun, and most
    pronounced for the high mass loss case in the lower panel, is due
    to a pile up of masses at the transition mass, $M_{\rm WN-WC}$,
    where the Yoon mass loss rate abruptly changes slope. See inset
    plots and \Fig{mdot} and \Fig{calibrate}. The sharpness of these
    peaks is artificial since the mass loss rate does not truly change
    slope abruptly and the observed black holes will not all come from
    progenitors with a single metallicity. Note the lack of any peaks
    in the 15 to 35 \Msun \ mass range for $Z_{\rm init}>0.3\ \Zsun$
    for the standard Yoon rate, and at $Z_{\rm init}>0.1\ \Zsun$ for
    twice the Yoon rate.
    \lFig{zdep}}
\end{figure}

% fig 8 - BH masses IMF dependence
\begin{figure}
\includegraphics[width=0.48\textwidth]{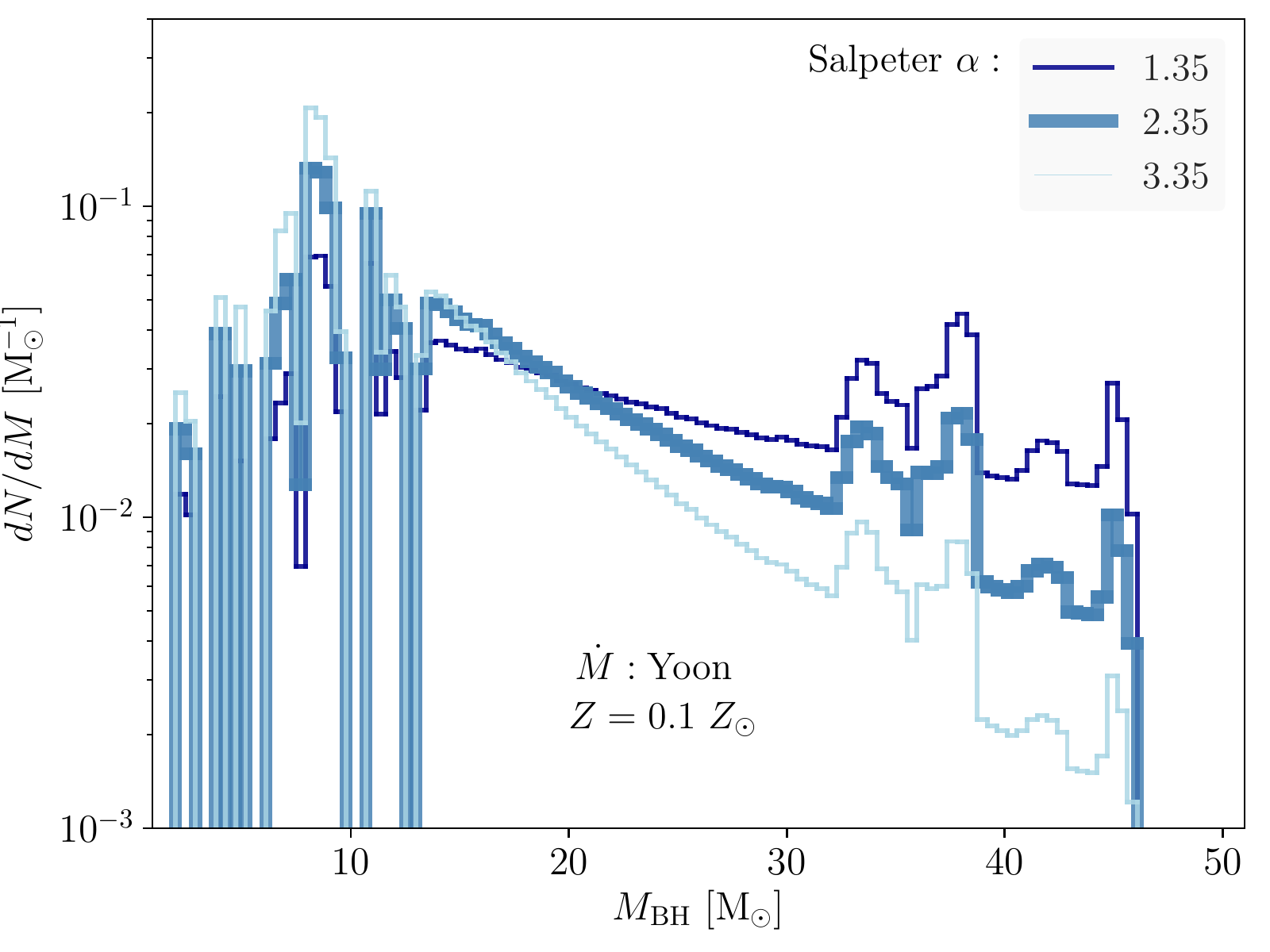}
\includegraphics[width=0.48\textwidth]{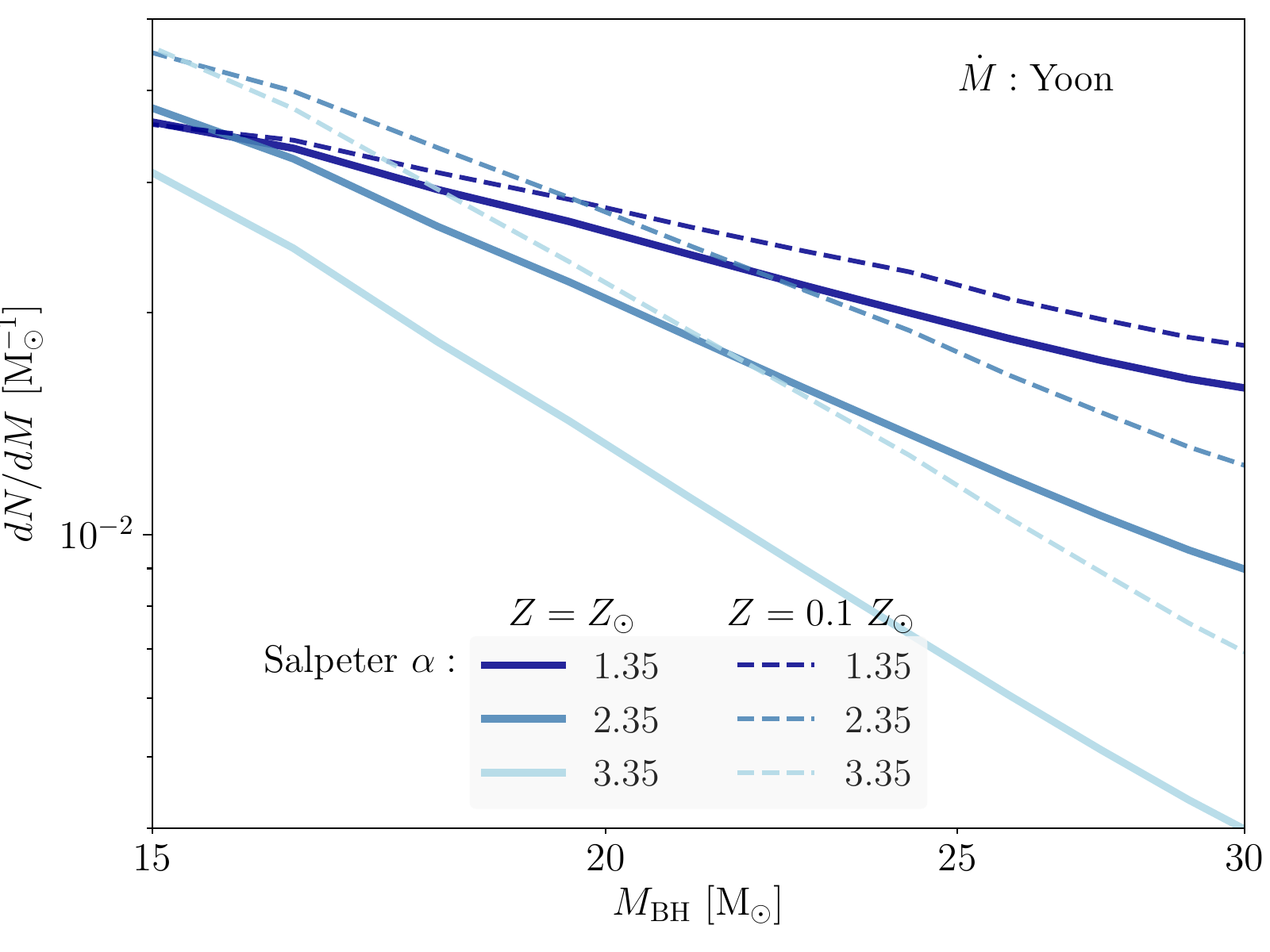}
\caption{The top panel shows the sensitivity of the black hole birth
  function to the assumed initial mass function for main sequence
  stars in close binaries for the standard Yoon mass-loss rate and
  several choices for the Salpeter power law exponent, $\alpha$. The
  value $\alpha$ = 2.35 is standard. The lower panel shows the
  dependence of the slope on both $\alpha$ and metallicity for the
  standard Yoon mass-loss prescription for a region of mass where all
  presupernovae are assumed to collapse to black holes with no mass
  loss except for what neutrinos carry away.
  \lFig{imf}}
\end{figure}

The relatively small values for the median black hole mass using the
Sander et al. mass-loss rate at solar metallicity are a consequence of
the pile up in presupernova masses around 8 -- 10 \Msun, resulting
from the very non--linear decrease in the mass-loss rate at around a
luminosity of 10$^{5.2}$ \Lsun\ (\Fig{mdot} and \Tab{hetable}). At lower
metallicity the dip in mass loss occurs at higher mass and also less
mass is lost, so the effect on the median mass is not so great. 

The median black hole masses when the entire mass range of initial
helium core masses, 2.5 to 150 \Msun, is sampled are substantially
larger than frequently cited by observers for binary X--ray
sources. \citet{Oze10} gives a mean mass of $7.8 \pm
1.2$\,\Msun. A similar value was found, in the case of a single
  Gaussian fit, by \citet{Far11}.  This lower value could reflect a
cut off in the production of very high mass stars coupled with a high
mass-loss rate, or observational bias. X--ray binaries containing
black holes above 20 \Msun\ (resulting from initial helium core masses
over 40 \Msun\ and main sequence masses over 80\Msun) may be rare,
short lived, and difficult to detect. \citet{Wyr16} have studied
  13 candidate compact objects using microlensing. The maximum mass
  black hole they observed had a mass of about 9 \Msun \ plus or minus
  a factor of two. Given the sparse sample and large error bar, this
  is consistent with our predictions. They also found no evidence for
  a mass gap between neutron stars and black holes. LIGO/Virgo has
turned up a large sample of much heavier black holes
\citep{Abb19}. This might be regarded as a more complete census of the
stellar graveyard for very high mass stars, since the black holes that
are merging might not be easily detected any other way.

The fraction of core collapses that produce black holes is sensitive to
the mass-loss rate employed. For a bigger mass-loss rate, a black hole
of given mass comes from a higher mass, rarer main sequence star. For
the full range of helium core masses considered, our computed fraction
ranges from 9\% to 32\% (\Tab{bhtbl}), which is consistent with a
prior estimate by \citet{Koc15} of 9\% to 39\%. When the sampling
interval is restricted to initial helium core masses below 40 \Msun
\ however, our fraction is smaller, as little as 4\% for Yoon mass-loss
rates with $f_{\rm WR}$ = 2. The maximum limit, regardless of metallicity
is 0.30.

An important feature in our results is the nearly constant slope of
the black hole birth function roughly between 15 and 35 \Msun. Above
35 \Msun, the distribution is altered by the PPISN, while below 15
\Msun\ stars frequently produce supernovae, and thus the distribution
is highly modular due to the varying outcomes of the stellar
collapse. In between lies a prominent range where the black hole mass
is solely determined by the presupernova mass, minus the small
correction for neutrino emission. If the mass-loss rate has no rapid
change in slope in this mass range, the birth function will be smooth
and if the dependence on mass nearly linear, it will mirror the
stellar IMF. For solar metallicity, the Vink and Yoon mass-loss rates
satisfy these criteria and the birth function in \Fig{zdep} is
smooth. For lower metallicities, the differing dependence on $Z_{\rm init}$ of
the rates for WN and WC mass loss cause a spike whose location
migrates with the changing $Z_{\rm init}$. 

The spike is due to a ``flat spot'' in the presupernova mass resulting
from the transition from WN to WC mass loss (\Fig{calibrate}).  Its
sharpness and location is probably an artifact of the empirical
mass-loss description for the two cases, which lead to a significant
jump between the WN and WC mass-loss when evaluated at lower
metallicities.  Such a significant jump in mass-loss rate between the
WN and WC stage is probably not realistic. In any case, the observed
distribution would be smoother as it would never include just one
metallicity, but a range. The \citet{San19a} rate also exhibits a similar
spike at low metallicity, but for a different reason. For 0.1 \Zsun,
the mass-loss rate is very non--linear for a luminosity of 10$^{5.8}$
\Lsun, corresponding to a mass near 20 \Msun \ (\Fig{mdot} and
\Tab{hetable}). This ledge has a physical basis, namely the strong
dependence of the Sander et al. rate on the Eddington Gamma--factor,
$\Gamma_e$, essentially the ratio of acceleration due to Thomson
scattering and gravity \citep{Vin06,San15}. A signature in the black
hole IMF would thus have interesting implications for the theory of
stellar mass loss.

The fact that, for solar metallicity at least, the slope of the birth
function between 15 and 35 \Msun\ in \Fig{bh} is nearly constant for
other choices of mass-loss rate reflects the near linear relation
between $\log \dot M$ and $M$ in that mass range (see \Fig{mdot} and 
\Tab{hetable}). Due to the Eddington limit, the lifetime of very 
massive stars is nearly constant at $\tau \approx$ 300,000 years, so 
the presupernova mass is almost a constant fraction of the initial 
helium core mass.
\begin{equation}
\Delta M/M \approx \ 1 \ - \ e^{-k \tau}
\end{equation}
where $k$ is positive and defined by $dM/dt = -k M$. Since all stars
in this mass range are assumed to collapse to black holes with
gravitational masses very nearly equal to presupernova masses, this
means that the remnant mass is a constant times the ZAMS mass, though
the value of that constant will vary with metallicity and mass-loss
rate.  Consequently, the slope of the birth function is mostly
sensitive to the IMF for massive stars (\Fig{imf}). Measuring the BH
birth function in this mass range with LIGO/Virgo offers an
opportunity to determine this important quantity. The corresponding
mass range sampled on the main sequence depends on the mass-loss rate
employed and can be read from \Fig{remsol} for the initial helium star
mass to be used in \eq{mzams} and \eq{mzamsh} For solar metallicity
and standard Yoon loss rates, the main sequence mass range sampled
would be approximately 70 to 150 \Msun. Sparse data currently exists
for main sequence stars with these masses due to their small birth
numbers and short lifetimes.

In the more general case that $dM/dt = - k M^\nu$, with $\nu$ not equal
to 1, the presupernova mass for very massive stars with constant lifetime
$\tau$ is
\begin{equation}
M_{\rm preSN} \ = \ \left( k \tau (\nu - 1) + M_{\rm He,i}^{1-\nu} \right)^{\frac{1}{1-\nu}}.
\end{equation}
For $\nu > 1$ this implies an upper limit to the mass of black holes of
\begin{equation}
M_{\rm max} \ = \ \left( k \tau (\nu
-1) \right)^{\frac{1}{1-\nu}}.
\end{equation}
For $\nu < 1$ there is no upper bound and the final black hole mass
continues to increase monotonically so long as helium cores of
arbitrarily high mass are produced. For the mass-loss rates considered
here, $\nu$ above 100 \Msun\ is so close to 1 that this upper limit is
much larger than the range considered here. For larger mass-loss rates
and larger $\nu$ though, this limit could be important. For example,
the mass dependent mass-loss rate of \citet{Lan89}, $dM/dt\approx
- 10^{-7} (M/\Msun)^{2.5}$ \Msun\ y$^{-1}$, gives $M_{\rm max}$= 8
\Msun. For solar metallicity, there would be no black holes with
greater mass, no matter how big the masses on the main sequence were.

\section{Discussion and Conclusions}
\lSect{conclude}

Using simple approximations to binary evolution and supernova
explosions, the birth functions for neutron stars and black holes have
been calculated for a variety of metallicities and modern mass-loss
rates. The evolution captures the essential nature of a helium star
that shrinks due to mass loss after losing its envelope, rather than a
helium core growing in a single star. The explosion model assumes that
the outcome, at least in terms of remnant mass, is determined by the
presupernova core mass independent of metallicity or interior
structure and composition. The presupernova mass is determined by
integration of the mass-loss equation, not by computing actual stellar
models. Within this framework a large set of results can be quickly
computed and general trends determined.

The results (Tables 3 and 4; Figs. 5, 6, and 7) show patterns that, to
good approximation, are independent of mass-loss rate and
metallicity. Neutron stars of all gravitational masses from 1.23 to
the maximum mass, here assumed to be 2.3 \Msun, are produced. The
distribution has a central peak near 1.35 \Msun, with rare events
producing much heavier neutron stars by fallback \citep[see
  also][]{Ert20}. Increased mass loss decreases the proportion of the
most massive members (\Fig{ns}).

The black hole birth function has four parts (\Fig{bh}). A low mass
tail below 5 \Msun, produced entirely by fallback in explosions that
initially make a neutron star, is followed by a broad distribution
resulting from both direct implosions and explosions with massive
fallback.  This distribution has a pronounced peak around 8 -- 9
\Msun. Its irregularity from 5 -- 12 \Msun\ reflects the mixed
distribution of stars that explode leaving neutron stars and implode
into black holes.  There is a hint of a gap around 10 \Msun\ due a
local minimum in the core compactness parameter for that presupernova
mass \citep{Woo19,Ert20}. The location of this gap may depend on
uncertainties in presupernova evolution like the
$^{12}$C($\alpha,\gamma)^{16}$O reaction rate. Above 12 \Msun, in the
present prescription all presupernova cores collapse promptly to black
holes and the distribution reflects the initial mass function for the
original stars modulated by any rapid non--linear behavior of the
mass-loss rate. Non--linearity can result from an abrupt transition
between WN and WC mass-loss formulae \citep{Yoo17}, or a rapid cut off
below a certain Eddington factor \citep{San19a}. The mass where the
inflection occurs is sensitive to the mass-loss rate and metallicity
(\Fig{zdep}) and might be spread out by observations that span a range
in metallicities. Without this structure, which is weak or absent in
the case of smooth \citep{Vin17} or small mass loss, the slope of the
black hole birth function mirrors that of the stellar IMF between
about 15 and 35 \Msun\ (\Fig{imf}). From 35 \Msun \ to 46 \Msun, there
is a pile up of PPISN remnants. More massive PPISN eject more mass so
there is a peak around 38 \Msun. The location of the peak and the
cutoff mass will be sensitive to the rate for
$^{12}$C($\alpha,\gamma)^{16}$O. All birth functions were calculated
assuming the continuation of a Salpeter--like IMF to helium core
masses as large as 150 \Msun. This corresponds to a ZAMS mass of about
300 \Msun. Any sharp variation in the IMF for lower masses would
imprint structure in the black hole birth function.

While our survey has used a range of mass-loss rates from the recent
literature, observational constraints and spectral analysis of Wolf-Rayet
(WR) stars point towards high mass-loss rates such as described by the
empirically motivated recipe of Yoon (2017). The number ratio of
observed Type WC and WN stars and the ejected masses of Type Ic supernovae
require that sufficient mass loss occurs, especially at solar
metallicity, to frequently uncover the carbon--oxygen core
\citep{Con76} before the star dies.

\citet{Neu11} report a ratio of massive WC to WN stars that ranges
from 58\% at solar metallicity to 9\% for the Small Magellanic Cloud
(SMC, roughly \Zsun/5). \citet{Neu19} give similar values, $0.83
  \pm 0.10$ for the Milky Way and $0.09 \pm 0.09$ for the
  SMC. \citet{Ros15} give a range from $0.51 \pm 0.08$ to $0.40 \pm
  0.16$ for the inner and outer Milky Way respectively. \Tab{wcwn}
shows our corresponding numbers for the standard Yoon, twice the Yoon,
and Sander et al. rates, each using \Eq{WN-WC} for the transition
mass. The theoretical fractions have been evaluated using the IMF
($\alpha=2.35$) weighted average of the time spent by each set of
models as a WC or WN star.

The model results depend upon the mass or luminosity range one
  ascribes to stars that are, observationally, Wolf-Rayet stars. This
  range depends upon metallicity. For example, according to
  \citet{She20}, only helium stars above $\log(L/\Lsun) \approx 4.9$
  will be, spectroscopically, classical WR stars in our present Galaxy
  \citep[see also][]{San19b}.  For the Large and Small Magellanic
  Clouds, \citet{She20} give larger, $\log(L/\Lsun)$ = 5.25 and 5.6
  \citep[see also][]{Hai14}. By this criterion, the rest of our helium
  stars would be ``stripped stars'', but not WR stars.  We thus
  evaluated our average WC/WN ratios for several cut-offs, three given
  by initial helium core mass and one by the luminosity of the star as
  it evolved. The mass cutoffs considered were 2.5 \Msun\ (initial
  $\log(L/\Lsun) = 3.76$); 4 \Msun\ (initial $\log(L/\Lsun) = 4.32$),
  and, for the two metallicities, either 7 \Msun\ (initial
  $\log(L/\Lsun) = 4.90$) or 16 \Msun \ (initial $\log(L/\Lsun) =
  5.60$). In the fourth case each star was only considered to be a WR
  star during the time its luminosity exceeded the threshold given by
  \cite{She20}. In this, and the higher mass cutoff cases, a much
  smaller fraction of stripped stars are WR stars. Most of the ones
  excluded are WN stars, so the WC/WN ratio is increased.

Many possible sources of error enter the comparison in \Tab{wcwn}. Not
all the observed WR stars were stripped by binary mass exchange. The
definition of a transition mass (eq.\Eqref{WN-WC}) is approximate and
arbitrary. The observed lower luminosity for classical WR stars is
approximate. If the cutoff is reduced from $\log(L/\Lsun) = 5.6$ to
5.25 in the case of the SMC, the ratio of WN/WC stars for our models
drops from 0.22 to 0.12 for the standard Yoon mass loss rates. More
realistic studies show good agreement, for the mass loss rates they
assume \citep{Van07,Eld17}. Still, our simple results suggest that
mass loss rates near those given by \citet{Yoo17}, with a multiplier
$f_{\rm WR}$ between 1 and 2, are favored over smaller ones. Following
similar arguments, \citet{Yoo17} himself suggested 1.58. The
\citet{Vin17} and \citet{San19a} rates produce too few WC stars.
  
A similar conclusion comes from considering SN Ib and SN Ic progenitor
masses.  Using the Yoon rates with $f_{\rm WR}$ = 1 gives a critical
presupernova mass separating WN progenitors from WC progenitors near
7.0 \Msun \ \citep{Woo19}. Increasing the rate by a factor of 1.5
reduces that critical mass to 4.9 \Msun, and a new calculation using
the KEPLER code shows that using $f_{\rm WR}$ = 2 lowers the critical
mass to 3.9. Assuming 1.5 \Msun\ is left in the neutron star, this
implies ejected masses of 5.5, 3.4, and 2.4 \Msun\ respectively.
Measured average ejected masses for Type Ic supernovae are near 2.2
\Msun\ \citep{Pre19}, suggesting $f_{\rm WR}$ = 2 might be
  appropriate.  Again there are caveats.  It is not clear
  that the production of a Type Ic supernova requires the complete
  loss of the helium shell \citep{Des12}.

\begin{deluxetable*}{lccccr}
\tablecaption{WC/WN ratios}
\tablehead{
          \colhead{$Z$}
        & \colhead{Observed}
        & \colhead{Yoon}
        & \colhead{Yoon$\times$2}
        & \colhead{Sander et al.}
        & \colhead{Cut-off}
          }
\startdata

\multirow{4}{*}{\Zsun}   & \multirow{4}{*}{0.5 -- 0.8} 
  & 0.05 & 0.15 & 0.02 & $M_{\rm He,i}>2.5\ \Msun$ \\
& & 0.12 & 0.42 & 0.06 & $M_{\rm He,i}>4.0\ \Msun$ \\
& & 0.33 & 1.74 & 0.14 & $M_{\rm He,i}>7.0\ \Msun$ \\
& & 0.49 & 1.30 & 0.15 & $\log L_{\rm He}  >4.9\ \Lsun$ \hspace{-0.3mm} \\
\\
\multirow{4}{*}{\Zsun/5} & \multirow{4}{*}{$\sim$0.1}
  & 0.01 & 0.03 & ---  & $M_{\rm He,i}>2.5\ \Msun$ \\
& & 0.02 & 0.08 & ---  & $M_{\rm He,i}>4.0\ \Msun$ \\
& & 0.20 & 0.98 & ---  & $M_{\rm He,i}>16\ \Msun$ \hspace{-0.5mm}\\
& & 0.26 & 0.99 & ---  & $\log L_{\rm He}  >5.6\ \Lsun$ \hspace{-2mm}
\enddata
\tablecomments{``Observed'' values are from \citet{Neu11}. The SMC is
  taken to be 0.2 \Zsun. No fits for or interpolation formulae for
  0.2 \Zsun \ were given by \citet{San19a}. WC/WN ratios obtained
  using the rates of \citet{Vin17} were much smaller.} \lTab{wcwn}
\end{deluxetable*}

Using Yoon's rates with $f_{\rm WR}$ = 2 gives a median neutron star
mass at solar metallicity of 1.32 \Msun \ (1.35 \Msun \ if the
Lattimer and Prakash (2001) correction for neutrino losses is employed
instead of the value calculated by Ertl et al (2020)), and a median
black hole mass of 7.9 \Msun, for initial helium cores with smaller
initial mass than 40 \Msun\ (ZAMS mass less than 80 \Msun). Given the
small fraction of black holes predicted by this cutoff (\Tab{bhtbl})
and the small maximum black hole mass implied (11.1 \Msun \ from a
presupernova of 11.7 \Msun; \Fig{remsol} and \Tab{death}), it may not
be realistic to truncate the maximum initial helium core mass at such
a low value if the mass-loss rate is this large. The median
  black hole mass including all initial helium star masses up to 150
  \Msun \ is 13.2 \Msun. In all cases, the maximum presupernova mass
  to collapse and leave a remnant is 60 \Msun \ and the maximum black
  hole mass, following pulsational activity, is 46 \Msun
  \ \citep[][\Fig{bh}]{Woo19}.

The formalism developed here is simple and easily applicable to more
realistic descriptions of binary evolution
\citep[e.g.,][]{Dom12,Fry12,Vig18,Spe15,Eld17}.  Once the helium
  core is uncovered the mass-loss equation is easily integrated and
  the remnant mass determined using Tables 1 and 2. Indeed, our
results are not only applicable to binary mass exchange. By adjusting
the lifetime in \Tab{hetable} to reflect the actual duration of the
Wolf-Rayet phase, presupernova masses can also be calculated for
single stars, and stars that experience chemically homogeneous
evolution \citep[e.g.,][]{deM16}.  Unless that adjustment is large,
the results presented here, nominally for close binaries, should be
robust. The main sequence masses used in weighting the contributions
in an IMF average would also need to be adjusted as in \eq{mzams} and
\eq{mzamsh} or their equivalents.

While this paper is about simple one-dimensional models, it is
possible to make some inferences about the rotation rates for the
black holes that are produced. This is because the final angular
momentum is very sensitive to the amount of mass lost by the helium
core during the stripped phase. The initial angular momentum of the
helium core can be quite large. A fit, to 5\% accuracy, to the radii
of the helium star models of \citet{Woo19} after they have burned only
1\% of their helium to carbon, is $R = 1.12 \times 10^{11} (M/30
\ \Msun)^{0.55}$ cm. The moment of inertia of a n = 3 polytrope, which
these stars resemble, is 0.0754 $MR^2$ \citep{Cri15}, though a
slightly better fit to the models will be used here, 0.09 $MR^2$. If
the helium star initially rotates with surface velocity equal to a
fraction $f_{\rm Kep}$ of Keplerian, $v_{\rm Kep} = (GM/R)^{1/2}$,
then the initial angular momentum will be
\begin{equation}
J_{\rm init} \ = \ 3.7 \times 10^{52} \ \left(\frac{f_{\rm Kep}}{0.33}\right)
\left(\frac{M}{30 \ \Msun}\right)^{1.775} \ {\rm erg \ s}.
\end{equation}
The Kerr parameter for the initial helium star, on the other hand, is
\begin{equation}
a_{\rm init} \ = \frac{J c}{G M^2} = 4.7 \ \left(\frac{f_{\rm Kep}}{0.33}\right) \left(\frac{M}{30
  \ \Msun}\right)^{0.225}.
\end{equation}
For rapid, but not extreme rotation the helium core can begin its life
with sufficient angular momentum to make a Kerr black hole. $f_{\rm
  Kep} = 0.33$ corresponds to a ratio of centrifugal force to gravity
of 10\%, which might be achieved by chemically homogeneous evolution
\citep{Woo06}.

As the star loses mass though, it loses angular momentum. About 80\%
of the angular momentum in the initial rigidly rotating star is
contained in the outer half of its mass. This fraction is amplified
appreciably by transport, especially by magnetic torques, as the star
burns helium. In practice, no helium star that begins with reasonable
rotation and loses half or more of its mass will produce a Kerr hole.
This includes all the models presented here that make black holes
using mass loss rates from \citet{Yoo17} with $f_{\rm WR} = 1$ or 2 at
solar metallicity. In fact, following transport in an actual KEPLER
model using the physics described in \citet{Woo06} for a helium core
with initial mass 60 \Msun \ and final mass 30 \Msun \ ($f_{\rm WR}$ =
1; Yoon rates) shows that losing half the mass reduces the angular
momentum by a factor close to 100. Approximately the same reduction is
also observed in a 30 \Msun \ star that loses half its mass. One thus
expects Kerr parameters $a \ltaprx 0.1$ for the black holes resulting,
even for stars with rapid initial rotation, from use of the Yoon rates
at solar metallicity. The Kerr parameter would be substantially larger
though at lower metallicity, especially if the rates of Vink or Sander
et al. are employed (\Fig{remsol}). Had the same 30 \Msun \ model lost
only 5 \Msun \ instead of 15 \Msun, it would have made a 25 \Msun
\ black hole with a = 1.

These qualitative arguments agree with the results of more detailed
studies \citep[e.g.][]{Bel20}. Unless the mass loss is small, the spins of
merging massive binary black holes are likely to be small.

\section*{Acknowledgments}

We thank the anonymous referee for many useful suggestions and 
references that helped to improve this work. The authors acknowledge 
extensive, educational correspondence with Andreas Sander on the topic 
of mass loss in massive helium stars. He, Jorick Vink and Sung-Chul 
Yoon made numerous suggestions that resulted in an improved manuscript. 
We also thank Carolyn Raithel for providing observationally inferred 
compact object mass distribution data. This work has been partly 
supported by NASA NNX14AH34G. TS was supported by NASA through a NASA 
Hubble Fellowship grant \#60065868 awarded by the Space Telescope 
Science Institute, which is operated by the Association of Universities 
for Research in Astronomy, Inc., for NASA, under contract NAS5-26555. 
At Garching, funding by the European Research Council through Grant 
ERC-AdG No.~341157-COCO2CASA and by the Deutsche Forschungsgemeinschaft 
(DFG, German Research Foundation) through Sonderforschungsbereich 
(Collaborative Research Centre) SFB-1258 ``Neutrinos and Dark Matter 
in Astro- and Particle Physics (NDM)'' and under Germany's Excellence 
Strategy through Cluster of Excellence ORIGINS (EXC-2094)---390783311 
is acknowledged.

\end{document}